\documentclass[a4paper, 11pt]{article}
\usepackage{aapreprint}
\usepackage{amsmath}
\usepackage{appendix}
\usepackage{graphicx}
\usepackage{subfigure}
\usepackage{amssymb}
\usepackage{xcolor}
\usepackage{multirow}
\usepackage{cancel}
\usepackage{color}
\usepackage{caption}
\usepackage{listings}
\usepackage{xcolor}
\usepackage{float}
\usepackage{slashed}
\usepackage{amsmath}
\usepackage{braket}
\usepackage{quantikz}
\usepackage{tikz}
\usetikzlibrary{quantikz2}
\usepackage{graphicx}
\usepackage{booktabs}
\usepackage{float}
\usepackage{amsmath}
\usepackage{amssymb}
\usepackage[utf8]{inputenc}
\usepackage{hyperref}
\usepackage{color}
\usepackage{slashed}
\usepackage[absolute]{textpos}
\usepackage{changes}
\allowdisplaybreaks

\newcommand{\SU}{\mathrm{SU}}
\newcommand{\Uone}{\mathrm{U}}
\newcommand{\dd}{\mathrm{d}}
\newcommand{\alphas}{\alpha_s}
\newcommand{\alphad}{\alpha_D}
\newcommand{\gammad}{\gamma_D}
\newcommand{\Mq}{M_q}
\newcommand{\LambdaD}{\Lambda_D}
\newcommand{\Qmin}{Q_{\min}}
\newcommand{\Qmax}{Q_{\max}}
\newcommand{\Nc}{N_C}
\newcommand{\Nf}{N_F}

\begin{document}

\title{Quantum Veto Parton Shower Algorithm in the Conformal Window}

\author[1]{Nongqiu Cai,}
\emailAdd{cnq4work@outlook.com}

\author[1]{Xu Deng,}
\emailAdd{xud119265@gmail.com}

\author[1,*]{Jinmian Li,\note[*]{Corresponding author.}}
\emailAdd{jmli@scu.edu.cn}

\author[1,2]{Bingwei Long,}
\emailAdd{bingwei@scu.edu.cn}
\affiliation[1]{College of Physics, Sichuan University, Chengdu 610065, China}
\affiliation[2]{Southern Center for Nuclear-Science Theory (SCNT), Institute of Modern Physics, Chinese Academy of Sciences, Huizhou 516000, Guangdong Province, China}

\author[3,4,*]{Shuo Yang}
\emailAdd{shuoyang@lnnu.edu.cn}
\affiliation[3]{School of Physics and Electronic Technology, Liaoning Normal University, Dalian, 116029, China}
\affiliation[4]{Center for Theoretical and Experimental High Energy Physics, Liaoning Normal University, Dalian, 116029, China}

\abstract{We develop a quantum veto parton shower algorithm for a confining dark sector with gauge symmetry $\mathrm{SU}(3)_D \times \mathrm{U}(1)_D$ operating in the conformal window ($N_C = 3$, $N_F = 12$). The non-Abelian gauge coupling exhibits two-loop running toward an infrared fixed point before decoupling at a finite dark-quark mass threshold, while the dark quarks carry non-degenerate $\mathrm{U}(1)_D$ charges with unitary flavor mixing between the interaction and mass bases. 
Shower evolution proceeds coherently in the interaction basis, with states prepared and measured in the mass basis. 
Compared with the classical Markovian shower, the flavor-coherent quantum simulation produces a significantly enhanced zero-emission rate, and a characteristic U-shaped mass-eigenstate spectrum. 
Furthermore, energy-energy correlators exhibit approximate linearity in the double-logarithmic coordinates over an intermediate angular range, suggesting power-law scaling in that region.
This framework demonstrates the capability of quantum algorithms to capture multi-flavor quantum interference in strongly interacting hidden sectors.}

\maketitle

\newpage
\section{Introduction} 
A compelling class of solutions to the nature of dark matter (DM) and the gauge hierarchy problem of the Standard Model (SM) postulates the existence of a ``Hidden Valley" or a dark sector~\cite{Strassler:2006im,Han:2007ae,Gori:2022vri}.
These frameworks typically introduce new degrees of freedom that are uncharged under the SM gauge group but interact through their own dark gauge symmetries. 
The phenomenology of these hidden sectors is remarkably rich, especially when the new dark forces are confining and weakly connected to the visible sector. 
If the confinement scale of the dark sector is low compared to the production energy scale at high-energy colliders, energetic dark partons will undergo extensive collinear splitting and subsequent hadronization, producing a collimated spray of dark hadrons~\cite{Cohen:2017pzm,Cohen:2020afv,Knapen:2021eip,Albouy:2022cin}. 
Depending on the spectrum, lifetime, and decay modes of these composite dark states, the resulting collider signatures can manifest in a variety of spectacular ways. 
They can appear as emerging jets when the dark hadrons decay visibly after traveling some distance~\cite{Schwaller:2015gea,CMS:2018bvr,Carrasco:2023loy,ATLAS:2025lfx,Carrasco:2025bct}, as semi-visible jets when only a fraction of the dark hadrons decay into SM particles~\cite{Cohen:2015toa,Beauchesne:2017yhh,CMS:2021dzg,Liu:2024rbe,Buckley:2025hty,ATLAS:2025kuz,Li:2026rfq}, as displaced vertices if the dark hadrons are long-lived~\cite{Cheng:2021kjg,Born:2023vll,Cheng:2024hvq,Cheng:2024aco,Liebersbach:2024kzc,Liu:2025bbc,CMS:2025mym,Chen:2025btv,Borrello:2026bfr}, as soft uncluttered energy patterns (SUEPs) when the dark shower is diffuse and low-thrust~\cite{Harnik:2008ax,Knapen:2016hky,CMS:2024nca}, or even as stopping dark mesons that are captured in the detector material before decaying~\cite{Asadi:2025vfr}. 

While significant efforts have been dedicated to searching for these dark shower signatures, the majority of phenomenological studies predominantly assume the dark gauge group to be purely QCD-like, {\text{i.e.}} an $SU(N)$ theory with a small number of light flavors. 
Under this assumption, dark showers are conventionally simulated using Markov Chain Monte Carlo (MCMC) frameworks such as Pythia8~\cite{Carloni:2010tw,Carloni:2011kk,Bierlich:2022pfr} and Herwig7~\cite{Kulkarni:2024okx,Kim:2026mfk}. The GlueShower package~\cite{Curtin:2022tou,Batz:2023zef} is also developed for pure Yang-Mills glueball scenarios. 
In these approaches, the classical simulation of parton showers~\cite{Collins:1989bt,Hoche:2014rga,Nagy:2017ggp,Forshaw:2020wrq,Papaefstathiou:2024qlg} relies on probabilistic approximations based on Altarelli-Parisi splitting functions and the dipole/antenna formalism. These classical methods enforce momentum conservation and color coherence step-by-step, such procedures may drop the phase information required to track the full quantum effects. 

Meanwhile, the resulting collider signature of the dark sector can differ substantially from an ordinary QCD jet when the hidden gauge theory contains a large number of fermion flavors and lies near or inside the conformal window~\cite{Kulkarni:2025rsl}. 
A distinctive feature of conformal-window theories is their non-QCD like running coupling. For massless dark quarks, the gauge coupling approaches an infrared fixed point, while for finite quark masses the theory can leave the conformal regime as the scale approaches the quark masses, since mass effects become important in the parton shower evolution. 
Such behavior modifies the structure of the dark parton shower and requires a simulation framework that goes beyond the usual QCD-like approximation. 
As recently highlighted in Ref.~\cite{Kulkarni:2025rsl}, standard simulation tools lack the exact two-loop running coupling dynamics required to capture the flow to an infrared fixed point at large flavor numbers, and the corresponding classical veto algorithm requires substantial modification to model the correct Sudakov evolution. 
Furthermore, the \emph{large number of flavors} typically present in a conformal window amplifies quantum interference between different flavor histories, an effect that cannot be fully captured by a purely classical Markov-chain simulation. 

To overcome the theoretical limit of classical MCMC processes, there has been a recent surge of interest in utilizing quantum computing algorithms to simulate high-energy physics phenomena~\cite{Bauer:2022hpo,Fang:2024ple,Bloch:2026rrw}, with parton showering emerging as a highly promising application~\cite{Yamazaki:2025bfq}. 
Instead of relying on classical probabilities, quantum parton showers evolve the state vector directly, which allows the algorithm to capture quantum interference between different flavor components that share a common kinematic history~\cite{Bauer:2019qxa, Macaluso:2021ngq,Gustafson:2022dsq,Deliyannis:2022uyh}. 
Novel algorithm paradigms such as quantum walks have been proposed to efficiently sample emission histories~\cite{Bepari:2021kwv}, and specific phenomenological targets such as dark sector radiation have been investigated within a quantum simulation framework~\cite{Chigusa:2022act}. 
Recent advancements have addressed the challenge of incorporating exact kinematic recoil into quantum circuits. The quantum veto parton shower (QVPS)~\cite{Bauer:2023ujy} dynamically combines quantum-state vector evolution with a classical veto procedure. This hybrid framework is capable of tracking the branching kinematics and reconstructing physical events while preserving the quantum interference effects for multi-flavor processes. 
This makes QVPS an ideal starting point for simulating dark showers whose flavor structure cannot be captured by classical MCMC evolution. 
These quantum methods also open a path to resolving the long-standing issue of spin correlations inside a jet~\cite{Richardson:2018pvo,Hamilton:2021dyz,CMS:2025awh}, which are intrinsically quantum-mechanical and are treated only approximately in classical showers. 

In this work, we extend the QVPS framework to a finite-mass conformal-window dark sector with gauge structure $SU(3)_D\times U(1)_D$. We consider the specific case of $N_C=3$ and $N_F=12$, utilizing the exact two-loop running coupling for the non-Abelian sector to capture the physics of the infrared fixed point. 
To model the extensive internal flavor space, we introduce twelve interaction eigenstates carrying different $U(1)_D$ charges. 
The parton shower evolutions in both the $SU(3)$ and $U(1)_D$ sector are encoded directly into the quantum circuits. 
This construction provides an explicit circuit-level realization of quantum interference in conformal window dark showers. It delivers a practical benchmark for studying how non-QCD-like running, large flavor multiplicities, and flavor-dependent emissions alter the shower observables at future colliders. 

This paper is organized as follows. In Section~\ref{sec:framework} we define the dark-sector model, derive the two-loop running coupling in terms of the Lambert-$W$ function, introduce the flavor mixing matrix, and review the classical parton shower formalism including the Sudakov factor, the evolution variable, and the discrete splitting densities. Section~\ref{sec:qvps} presents the quantum veto parton shower algorithm: the history-space formalism, the hybrid classical-proposal--quantum-veto procedure, the qubit encoding and circuit design, and the analytic verification of the algorithm in simple cases. Our numerical results for the full shower are presented in Sec.~\ref{sec:results}. We conclude in Section~\ref{sec:conclusion}.

\section{Conformal-Window Dark Sector and Classical Parton Shower}
\label{sec:framework}

\subsection{Dark-sector model}

Let $\boldsymbol{\chi}=(\chi^{\mathrm{int}}_0,\ldots,\chi^{\mathrm{int}}_{11})^{T}$ denote the twelve Dirac dark quarks.
They span the interaction basis of the dark flavor space, and throughout this paper the superscript ``int'' marks interaction-basis quantities, while the $\chi_m$ ($m=0,\ldots,11$) without superscript refer to the mass-basis eigenstates.
The part of the Lagrangian relevant for the perturbative shower is
\begin{align}
  \mathcal{L}_{D}
  ={}&
  -\frac{1}{4}G_{\mu\nu}^{a}G_a^{\mu\nu}
  -\frac{1}{4}F_{\mu\nu}^{D}F_D^{\mu\nu}
  +\overline{\boldsymbol{\chi}}
   \left(i\slashed{\partial}-\Mq\mathbf{1}_{12}\right)
   \boldsymbol{\chi}
  \nonumber\\
  &+
  g_s\,
  \overline{\boldsymbol{\chi}}\gamma^\mu T^a
  \boldsymbol{\chi}\,G_\mu^a
  +
  g_D\,
  \overline{\boldsymbol{\chi}}\gamma^\mu
  \mathbf{Q}\boldsymbol{\chi}\,A_\mu^D ,
  \label{eq:dark-lagrangian}
\end{align}
where $G_\mu^a$ and $A_\mu^D$ are the $\SU(3)_D$ and $\Uone(1)_D$ gauge fields, respectively, with field-strength tensors $G_{\mu\nu}^a$ and $F_{\mu\nu}^D$.
The generators of $\SU(3)_D$ in the fundamental representation are $T^a$, with the dark strong coupling $g_s$ ($\alphas=g_s^2/4\pi$). Note that $g_s$ and $\alphas$ refer to the dark $\SU(3)_D$ interaction and are unrelated to the Standard Model strong coupling, which does not appear in this work.
The Abelian coupling is $g_D$ ($\alphad=g_D^2/4\pi$).
For simplicity, we assume all twelve dark quarks share a common mass $\Mq$, so that the mass term is proportional to the identity. The model predicts no preferred relation between the interaction basis and the basis in which the fermions are measured.  The mixing matrix used in this prototype is therefore an imposed benchmark ansatz rather than a matrix derived from a non-degenerate mass spectrum.

The charge matrix $\mathbf{Q}$ in the interaction basis is diagonal:
\begin{equation}
  \mathbf{Q}
  =
  \operatorname{diag}(Q_0,\ldots,Q_{11}),
  \qquad
  Q_k
  =
  \Qmin+k\frac{\Qmax-\Qmin}{\Nf-1},
  \qquad
  \Nf=12 .
  \label{eq:charge-matrix}
\end{equation}
In this setup, the charges are uniformly spaced between $\Qmin=0.5$ and $\Qmax=2.0$ (in units of the fundamental $\Uone(1)_D$ charge).
Consequently, the effective Abelian coupling of interaction eigenstate $k$ is
\begin{equation}
  \alpha_D^{(k)}=\alphad Q_k^2 .
  \label{eq:alpha-d-charge}
\end{equation}

\subsection{Two-loop running coupling and the Lambert-$W$ function}

For the $\SU(3)_D$ DM sector with $\Nc=3$ and $\Nf=12$, the renormalization-group equation for the gauge coupling~\cite{Kulkarni:2025rsl},
\begin{equation}
  \frac{\dd\alphas}{\dd\ln\mu^2}
  =
  -\beta_0\alphas^2-\beta_1\alphas^3+\mathcal{O}(\alphas^4),
  \label{eq:beta-rg}
\end{equation}
is governed by the one- and two-loop coefficients in the $\overline{\mathrm{MS}}$ scheme:
\begin{align}
  \beta_0
  &=
  \frac{1}{4\pi}
  \left(
    \frac{11}{3}C_A-\frac{4}{3}T_F\Nf
  \right),
  \nonumber\\
  \beta_1
  &=
  \frac{1}{(4\pi)^2}
  \left[
    \frac{34}{3}C_A^2
    -
    \left(
      \frac{20}{3}C_A+4C_F
    \right)T_F\Nf
  \right],
  \label{eq:beta-coefficients}
\end{align}
with the quadratic Casimir invariants $C_A=\Nc=3$, $C_F=(\Nc^2-1)/(2\Nc)=4/3$, and the Dynkin index $T_F=1/2$.
One obtains $\beta_0=3/(4\pi)>0$ and $\beta_1=-50/(4\pi)^2<0$.

In this case, the negative two-loop coefficient drives the running coupling toward an infrared fixed point~\cite{Caswell:1974gg,Banks:1981nn}:
\begin{equation}
  \alpha_*
  =
  -\frac{\beta_0}{\beta_1}
  =
  \frac{12\pi}{50}
  \approx 0.754 .
  \label{eq:alpha-star}
\end{equation}
This places the theory inside the conformal window, where the gauge coupling flows to a finite value in the infrared rather than diverging as in ordinary QCD.

The two-loop renormalization-group equation~\eqref{eq:beta-rg} can be solved exactly~\cite{Gardi:1998qr,Gardi:1998rf,Magradze:1999um,Prosperi:2006hx}.
Using $\alpha_*=-\beta_0/\beta_1$, the beta function $\beta(\alpha)\equiv\dd\alpha/\dd\ln\mu^2$ can be written in factorized form,
\begin{equation}
  \beta(\alpha)
  =
  \frac{\dd\alphas}{\dd\ln\mu^2}
  =
  -\beta_0\alphas^2-\beta_1\alphas^3
  =
  -\beta_0\alphas^2\!\left(1-\frac{\alphas}{\alpha_*}\right),
  \label{eq:beta-factorized}
\end{equation}
which makes the infrared fixed point at $\alpha_*$ manifest.
Integrating Eq.~\eqref{eq:beta-factorized} yields
\begin{equation}
  \ln\frac{\mu^2}{\LambdaD^2}
  =
  \frac{1}{\beta_0}
  \left[
    \frac{1}{\alphas}
    +
    \frac{1}{\alpha_*}
    \ln\!\left(\frac{\alpha_*-\alphas}{\alphas}\right)
  \right],
  \label{eq:implicit-solution}
\end{equation}
where the integration constant $\LambdaD$ is the characteristic scale around which the coupling transitions from the asymptotically free ultraviolet regime to the infrared conformal regime. At $\mu^2 = \LambdaD^2$, the coupling is fixed to the specific value $\alphas(\LambdaD^2)\approx 0.59$ according to Eq.~\ref{eq:implicit-solution}. 
In the infrared limit $\mu^2\ll\LambdaD^2$, the left-hand side is large and negative; the logarithmic term on the right then forces $\alphas\to\alpha_*$, reproducing the infrared fixed-point behavior.
Conversely, for $\mu^2\gg\LambdaD^2$ the right-hand side is dominated by $1/\alphas$, giving the one-loop running $\alphas\sim 1/(\beta_0\ln(\mu^2/\LambdaD^2))$ characteristic of asymptotic freedom.

Equation~\eqref{eq:implicit-solution} can be inverted analytically in terms of
the principal branch $W_0$ of the Lambert-$W$ function:
\begin{equation}
    \alphas(\mu^2)=
    \frac{\alpha_*}{1+W_0\!\left(x\right)},~
      x(\mu^2) \equiv
    e^{-1}
    \left(\frac{\mu^2}{\LambdaD^2}\right)^{\!\gamma},
    ~\gamma=-\frac{\beta_0^2}{\beta_1}=\frac{9}{50}.
  \label{eq:lambert}
\end{equation}
Here the Lambert-W function is defined as $W_0(x)e^{W_0(x)}=x$, and $\gamma$ controls the transition between
the infrared fixed point and the asymptotically free ultraviolet regime.
For every physical scale $\mu^2>0$, $x(\mu^2)>0$, the solution is real
and single-valued. Numerically, we evaluate the principal branch using
\texttt{scipy.special.lambertw} with $k=0$~\cite{Virtanen_2020},
taking its real part and enforcing a non-negative result only to guard against
floating-point artifacts. For $\mu^2\leq0$, the implementation returns the
infrared limiting value $\alpha_*$. The finite dark-quark mass $\Mq$
provides an exit from the conformal regime near the mass scale.

\begin{figure}[t]
\centering
\includegraphics[width=0.6\linewidth]{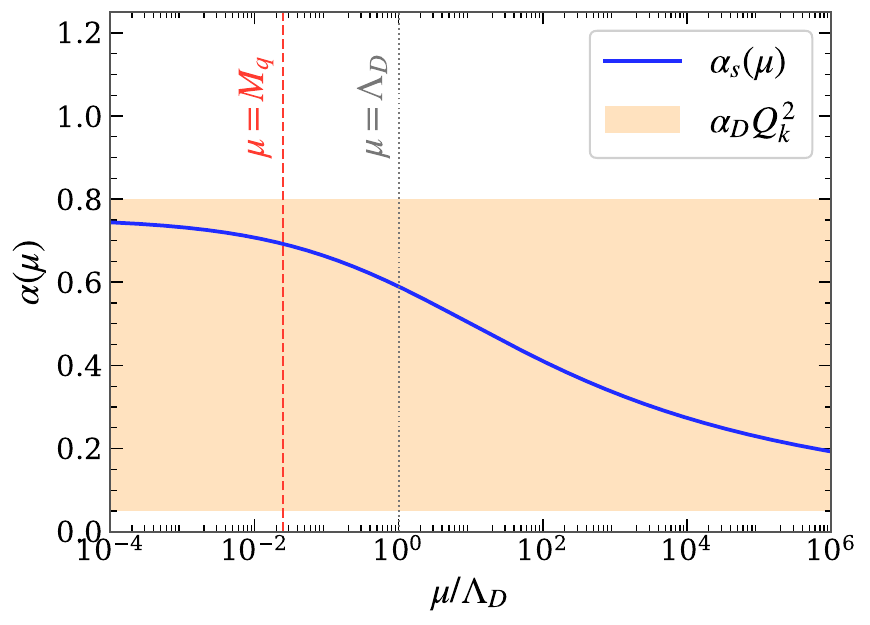}
\caption{Running couplings as a function of the scale ratio $\mu/\Lambda_D$. The solid blue curve depicts the exact two-loop $\mathrm{SU}(3)_D$ coupling $\alpha_s(\mu)$ with $N_C = 3$ and $N_F = 12$, flowing toward the infrared fixed point of Eq.~\eqref{eq:alpha-star}. The orange shaded band spans the effective $\mathrm{U}(1)_D$ couplings $\alpha_D Q_k^2$ for the twelve interaction eigenstates ($Q_k \in [0.5, 2.0]$ with $\alpha_D = 0.2$). Vertical dashed reference lines indicate the transition scale $\mu = \Lambda_D$ (grey) and the dark-quark mass scale $\mu = M_q$ (red).
\label{fig:coupling}}
\end{figure}

Figure~\ref{fig:coupling} illustrates the gauge coupling inputs across energy scales $\mu/\Lambda_D$. The non-Abelian coupling $\alpha_s(\mu)$ follows the exact two-loop conformal-window trajectory given in Eq.~\eqref{eq:lambert}, transitioning from the asymptotically free UV regime to the conformal fixed-point regime where it approaches the infrared fixed point $\alpha_*$ of Eq.~\eqref{eq:alpha-star}. In contrast, the Abelian coupling is non-running in our benchmark and forms a charge-dependent band $\alpha_D Q_k^2 \in [\alpha_D Q_{\min}^2, \alpha_D Q_{\max}^2]$ across the twelve interaction eigenstates. 
The vertical dashed lines mark the characteristic scale $\Lambda_D$ (grey) and the dark-quark mass threshold $\mu = M_q$ (red). In the physical shower evolution, the running is terminated at the mass scale $\mu \sim M_q$, below which the twelve dark flavors decouple from the renormalization-group flow and the evolution exits the conformal regime.

\subsection{Flavor bases and mixing matrix}
\label{sec:flavor-bases}

The interaction eigenstates
\begin{equation}
  \ket{k}_{\mathrm{int}},
  \qquad
  k=0,\ldots,11,
\end{equation}
form the basis in which the charge matrix and all gauge interactions are diagonal, and satisfy $\ket{k}_{\mathrm{int}}=\ket{\chi^{\mathrm{int}}_k}$.  The mass-basis states $\chi_m$ used for state preparation and final measurement are related to the interaction basis by a unitary transformation.

Because all dark quarks are assigned a common mass, the model does not predict any particular mixing between the interaction and mass bases. The transformation introduced here is a benchmark ansatz.  As a maximal-mixing benchmark, we first define the normalized twelve-point discrete Fourier transform,
\begin{equation}
  V_{mk}
  =
  \frac{1}{\sqrt{12}}
  \exp\left(\frac{2\pi i\,m k}{12}\right),
  \qquad
  m,k=0,\ldots,11 .
  \label{eq:dft-basis}
\end{equation}
All elements of this matrix have equal magnitude,
\begin{equation}
  \left|V_{mk}\right|^2
  =
  \frac{1}{12},
  \label{eq:dft-maximal-mixing}
\end{equation}
which makes this matrix a natural maximal-mixing benchmark for studying flavor-interference effects.

To vary the mixing strength continuously, we generalize this special choice to a one-parameter family of unitary transformations.  The Hermitian generator associated with the maximal-mixing matrix $V$ is defined using the principal matrix logarithm,
\begin{equation}
  H
  =
  -i\,\operatorname{Log}V .
  \label{eq:dft-generator}
\end{equation}
The general flavor-mixing transformation is then
\begin{equation}
  U(\xi)
  =
  \exp\left(i\xi H\right),
  \qquad
  0\leq\xi\leq1 .
  \label{eq:parameterized-mixing}
\end{equation}
This construction preserves unitarity throughout the interpolation,
\begin{equation}
  U^\dagger(\xi)U(\xi)
  =
  I_{12},
  \label{eq:mixing-unitarity}
\end{equation}
and has the endpoint values
\begin{equation}
  U(0)
  =
  I_{12},
  \qquad
  U(1)
  =
  V .
  \label{eq:mixing-endpoints}
\end{equation}
The lower endpoint corresponds to aligned interaction and mass bases, whereas the upper endpoint reproduces the maximal-mixing benchmark.  Intermediate parameter values provide a continuous interpolation between these limits.

For a general value of the mixing parameter, the basis states are related by
\begin{equation}
  \ket{m}_{\mathrm{mass}}
  =
  \sum_{k=0}^{11}
  U_{mk}^*(\xi)\ket{k}_{\mathrm{int}} .
  \label{eq:flavor-basis-transformation}
\end{equation}
As a benchmark choice, the initial hard process is taken to prepare the lowest-index mass-basis state, which is represented in the interaction basis as
\begin{equation}
  \ket{\chi_0(\xi)}
  =
  \ket{0}_{\mathrm{mass}}
  =
  \sum_{k=0}^{11}
  U_{0k}^*(\xi)\ket{k}_{\mathrm{int}} .
  \label{eq:parameterized-chi0-state}
\end{equation}
At the maximal-mixing endpoint used for the numerical results, this state becomes the uniform superposition
\begin{equation}
  \ket{\chi_0(1)}
  =
  \frac{1}{\sqrt{12}}
  \sum_{k=0}^{11}\ket{k}_{\mathrm{int}} .
  \label{eq:chi0-state}
\end{equation}

In the numerical construction of intermediate transformations, the generator is explicitly Hermitian-symmetrized,
\begin{equation}
  H
  \longrightarrow
  \frac{1}{2}\left(H+H^\dagger\right),
  \label{eq:generator-symmetrization}
\end{equation}
to remove floating-point noise introduced by the numerical evaluation of the matrix logarithm.

The charge-dependent veto constructed in Sec.~\ref{sec:qvps} rotates the emission qubit by a flavor-dependent angle, so that the relative flavor amplitudes of the evolved state $\sum_k c_k\ket{k}_{\mathrm{int}}$ are reweighted at every veto step even though no flavor-changing vertex is applied. The explicit form of this rotation is given in Eqs.~\eqref{eq:quantum-veto-action} and \eqref{eq:block-veto}.
At the end of each event, the mixing matrix $U(\xi)$ is applied to rotate the final state into the benchmark measurement basis.
If the coefficients $c_k$ remained equal, this rotation would project the state entirely onto the initial mass eigenstate $\ket{0}_{\mathrm{mass}}$, so only the non-uniformity of the $c_k$ generated by the flavor-dependent veto steps produces a nonzero overlap with the other mass eigenstates $m\neq0$.
The resulting mass-eigenstate distribution is the primary quantum interference observable studied in Sec.~\ref{sec:results}.

\subsection{Classical parton shower kinematics and Sudakov factor}

The parton shower is ordered in the Herwig-like evolution variable
$t^2\equiv\widetilde q^{\,2}$~\cite{Bahr:2008pv}, which decreases monotonically from the
hard starting scale $t_0^2$ to the infrared evolution-scale cutoff
$t_{\mathrm{cut}}^2$.  
Following Ref.~\cite{Bauer:2023ujy}, the evolution variable is
discretized according to
\begin{equation}
  t_{j+1}^2=r\,t_j^2,
  \qquad 0<r<1,
  \label{eq:discrete-virtuality}
\end{equation}
where $r$ is chosen sufficiently close to unity that the probability
of two resolved branchings in a single evolution interval is
negligible.  At a fixed step, the candidate scale $t_j^2$ is selected
before the energy fraction $z$ is sampled.  Once $z$ is known, the
physical off-shellness, transverse momentum, and daughter kinematics
are reconstructed from the sampled pair $(t_j^2,z)$.

For an accepted splitting $i\to j_1j_2$ at
$t_{\mathrm{split}}^2$, with $z$ carried by $j_1$, the parent line
ends at $t_{\mathrm{split}}^2$.  Angular ordering sets the maximum
subsequent evolution scales of the daughter lines to
\begin{equation}
  t_{j_1,\max}^2=z^2t_{\mathrm{split}}^2,
  \qquad
  t_{j_2,\max}^2=(1-z)^2t_{\mathrm{split}}^2.
  \label{eq:daughter-virtualities}
\end{equation}
The factors are squared because the Herwig relation is
$\widetilde q_{j_1,\max}=z\widetilde q_{\mathrm{split}}$ and
$\widetilde q_{j_2,\max}=(1-z)\widetilde q_{\mathrm{split}}$, whereas
the simulation stores the squared variable $t^2=\widetilde q^{\,2}$.

The present implementation imposes a common symmetric mass bound on
all enabled channels by requiring both daughter starting scales to be
at least $M_Q^2$ (infrared cut-off for dark photon and dark gluon).  This gives
\begin{equation}
  z_-(t^2)=\sqrt{\frac{M_Q^2}{t^2}},
  \qquad
  z_+(t^2)=1-z_-(t^2).
  \label{eq:z-limits}
\end{equation}
The interval closes at
$t^2=4M_Q^2$, where $z_-=z_+=1/2$. So the main shower uses
$t_{\mathrm{cut}}^2=4M_Q^2$ as its infrared evolution-scale cutoff.

The splitting kernels employed in the implementation are the standard unregularized Altarelli--Parisi functions:
\begin{align}
  P_{ff}(z) &= \frac{1+z^2}{1-z},
  \qquad
  P_{fA}(z) = z^2+(1-z)^2,
  \nonumber\\
  P_{gg}(z) &= \frac{z}{1-z}+\frac{1-z}{z}+z(1-z).
  \label{eq:splitting-kernels}
\end{align}
Here $P_{ff}$ governs the fermion$\to$fermion+gauge-boson splitting ($\chi\to\chi g$, $\chi\to\chi\gammad$), $P_{fA}$ describes the gauge-boson$\to$fermion-pair splitting ($g\to\chi\bar{\chi}$, $\gammad\to\chi\bar{\chi}$), and $P_{gg}$ is the triple-gluon vertex ($g\to gg$).
The colour and charge factors are not included in $P_{i\to a}$, but in the effective coupling $\alpha_{i\to a}$ defined below.

The differential probability for a branching $i\to a$ to occur at shower-evolution scale $t^2$ with energy fraction $z$ is encoded in the splitting density
\begin{equation}
  \frac{\dd\mathcal{P}_{i\to a}}{\dd\ln t^2\,\dd z}
  =
  \frac{\alpha_{i\to a}(\mu^2)}{2\pi}\,
  P_{i\to a}(z),
  \label{eq:diff-splitting}
\end{equation}
where $\alpha_{i\to a}$ denotes the effective coupling of the channel, evaluated at a renormalization scale set by the kinematics of the splitting, and includes the appropriate colour or charge factor of the vertex:
\begin{equation}
  \begin{aligned}
    \alpha_{\chi\to\chi g}&=C_F\alphas, &\quad
    \alpha_{g\to gg}&=C_A\alphas, &\quad
    \alpha_{g\to\chi\bar\chi}&=T_F\alphas, \\[2pt]
    \alpha_{\chi\to\chi\gammad}&=\alpha_D^{(k)}, &\quad
    \alpha_{\gammad\to\chi\bar\chi}&=\alpha_D^{(j)}, &\quad
  \end{aligned}
  \label{eq:effective-couplings}
\end{equation}
with $\alpha_D^{(k)}=\alphad Q_k^2$ from Eq.~\eqref{eq:alpha-d-charge} and $\alphas$ the dark $\SU(3)_D$ coupling.

For fermion radiation, the strong and dark-Abelian channels use
\begin{equation}
  \mu_{\chi\to\chi g}^{2}
  =\mu_{\chi\to\chi\gammad}^{2}
  =z(1-z)^2t^2.
  \label{eq:renorm-scale-ff}
\end{equation}
For the boson-to-fermion-pair channels and the triple-gluon channel,
the implementation uses the transverse-momentum-type scale
\begin{equation}
  \mu_{g\to\chi\bar\chi}^{2}
  =\mu_{\gammad\to\chi\bar\chi}^{2}
  =\mu_{g\to gg}^{2}
  =k_\perp^{2}
  =z^2(1-z)^2t^2.
  \label{eq:renorm-scale-fA}
\end{equation}
Here $\mu$ is the process-dependent renormalization scale~\cite{Bassetto:1983mvz,Amati:1980ch,Bauer:2023ujy} used in
the coupling evaluation and should not be confused with the global
shower-evolution variable $t^2$. 
However, the $\Uone(1)_D$ coupling $\alphad$ is taken to be
constant in the present model. Consequently, the renormalization scales
assigned to the $\chi\to\chi\gammad$ and
$\gammad\to\chi\bar{\chi}$ channels are not used in the subsequent
coupling evaluation; only the running $\SU(3)_D$ coupling $\alphas$ is
evaluated at its corresponding channel-dependent scale.

The central object of any parton shower is the Sudakov form factor, which gives the probability that no branching occurs while the parton evolves from the shower-evolution scale $t_2^2$ down to $t_1^2$ (with $t_1^2<t_2^2$):
\begin{equation}
  \Delta_i(t_1^2;t_2^2)
  =
  \exp\left[
    -\int_{t_1^2}^{t_2^2}\frac{\dd t^2}{t^2}
     \sum_a
     \int_{z_-(t^2)}^{z_+(t^2)}
     \dd z\,
     \frac{\dd\mathcal{P}_{i\to a}}{\dd\ln t^2\,\dd z}
  \right].
  \label{eq:sudakov}
\end{equation}
The sum over $a$ runs over all allowed branching channels for parton $i$.
It captures the resummation of the leading logarithmic contributions from soft and collinear radiation and is the essential building block that ensures unitarity of the parton shower.

The channel-by-channel forms of the differential probability of Eq.~\eqref{eq:diff-splitting} are collected in the vertex factors $\omega_{i\to a}(z, \mu^2) \equiv \dd\mathcal{P}_{i\to a}/(\dd\ln t^2\,\dd z)$:
\begin{equation}
\begin{aligned}
\omega_{\chi \to \chi g}(z, \mu^2) &= \frac{C_F \alpha_s(\mu^2)}{2\pi} P_{ff}(z), \\
\omega_{\chi \to \chi \gamma_D, k}(z) &= \frac{\alpha_D^{(k)}}{2\pi} P_{ff}(z), \\
\omega_{g \to gg}(z, \mu^2) &= \frac{C_A \alpha_s(\mu^2)}{2\pi} P_{gg}(z), \\
\omega_{g \to \chi\bar{\chi}, j}(z, \mu^2) &= \frac{T_F \alpha_s(\mu^2)}{2\pi} P_{fA}(z), \\
\omega_{\gamma_D \to \chi\bar{\chi}, j}(z) &= \frac{\alpha_D^{(j)}}{2\pi} P_{fA}(z),
\end{aligned}
\label{eq:differential_vertices}
\end{equation}
where $\mu^2$ takes the channel-dependent kinematic scale defined in Eqs.~\eqref{eq:renorm-scale-ff} and \eqref{eq:renorm-scale-fA}, and the effective couplings are those of Eq.~\eqref{eq:effective-couplings}.  Since $\alphad$ is constant in the present model, the two dark-photon vertices do not depend on $\mu^2$.
Notice that the gluon emission and self-interaction vertices ($\chi \to \chi g$, $g \to gg$) are flavor-independent, whereas the dark-photon vertices ($\chi \to \chi \gamma_D$, $\gamma_D \to \chi\bar{\chi}$) depend explicitly on the dark-quark charge $Q_k$.  For the $\gamma_D\to\chi\bar\chi$ channel the veto angle used in the algorithm is nevertheless flavor-independent, because it is built from the density summed over the twelve produced flavors, as described in Sec.~\ref{sec:qvps}.

For the discretized evolution with step size $\Delta \ln t^2 = -\ln r$, the target probability density in interval $[t^2_j, t^2_{j-1}]$ is directly related to the vertex factors by:
\begin{equation}
g_{i\to a}^{(j)}(t^2, z) = (-\ln r)\,\omega_{i\to a}(z, \mu^2),
\label{eq:discrete_densities}
\end{equation}
with the total pair-production densities summed over all $N_F = 12$ final flavors ($g_{g\to \chi\bar{\chi}}^{(j)} = \sum_j g_{g\to \chi\bar{\chi}, j}^{(j)}$ and $g_{\gamma_D \to \chi\bar{\chi}}^{(j)} = \sum_j g_{\gamma_D \to \chi\bar{\chi}, j}^{(j)}$).
These discrete densities provide the physical target distributions for the quantum veto algorithm discussed in the next section.

In the classical parton shower, one generates branchings by solving the equation $\Delta_i(t^2;t_{\mathrm{start}}^2) = \mathcal{R}$, where $\mathcal{R}\in[0,1]$ is a uniformly distributed random number.
Because the exact Sudakov factor is generally not analytically invertible, practical implementations employ the veto algorithm: one constructs an overestimate $\overline{g}_{ia}^{(j)}\ge g_{ia,k}^{(j)}$ that can be integrated and inverted analytically, proposes candidate branchings from the overestimate, and then accepts or vetoes each proposal with probability $g_{ia,k}^{(j)}/\overline{g}_{ia}^{(j)}$.
The hybrid quantum veto algorithm described in Sec.~\ref{sec:qvps} extends this classical procedure to the quantum domain by encoding the accept/reject decision into conditional quantum gates acting on the flavor register.

\section{Quantum Veto Parton Shower Algorithm and Circuit Realization}
\label{sec:qvps}

\subsection{Overview of the quantum veto approach}

The quantum veto parton shower~\cite{Bauer:2023ujy} is a hybrid algorithm in which the kinematic variables of each candidate branching are sampled on a classical computer, while the accept/reject decision is implemented as a quantum gate conditioned on the flavor state of the radiating parton.
At each discretized evolution-scale $t_j^2$, the classical side evaluates the overestimated splitting densities, selects a candidate parton $i$ and branching channel $a$, samples an energy fraction $z$, and computes the corresponding acceptance probability. The grid value $t_j^2$ is fixed before $z$ is sampled. The sampled $z$ is then used together with $t_j^2$ to reconstruct the physical branching kinematics.  
The quantum side then applies a controlled rotation to an emission qubit, whose measurement outcome determines whether the proposed branching is accepted.
This separation ensures that the full kinematic information is available for event reconstruction, while the flavor register itself is never measured until the end of the shower.  Flavor components that share the same kinematic history therefore remain coherent, whereas different histories are combined incoherently, as discussed in the history-space formalism below.

The flavor state evolves in the interaction basis where all charge-dependent vertices are diagonal.
Branchings with a flavor-independent veto angle ($\chi\to\chi g$, $g\to gg$, $g\to\chi\bar{\chi}$ and $\gammad\to\chi\bar{\chi}$) apply the same rotation angle to every flavor component and therefore preserve the relative flavor amplitudes.
Only the $\chi\to\chi\gammad$ channel has a flavor-dependent veto angle, applying different rotation angles to different charge eigenstates and thereby imprinting the charge structure onto the flavor register.
At the end of the shower, the final flavor state is rotated to the mass basis by the mass mixing matrix and measured, converting the accumulated amplitude differences into observable interference patterns.
The source code implementing the QVPS is available on Github \url{https://github.com/cai4work/Quantum-Parton-Shower-in-the-Conformal-Window}.

\subsection{Classical proposal and quantum veto}
The QVPS algorithm proceeds step by step on the discretized shower-evolution grid.
At each step $j$, for every active particle $i$ and channel $a$, an analytically integrable overestimate density
$\overline g_{ia}^{(j)}(t^2,z)$
is constructed such that it bounds the physical target density from above for all flavor components:
\begin{equation}
\overline g_{ia}^{(j)}(t^2,z)\;\ge\; g_{ia,k}^{(j)}(t^2,z)
\qquad\forall\,k.
\label{eq:overestimate-condition}
\end{equation}
The integrated overestimates are
\begin{equation}
  \overline G_{ia}^{(j)}(t^2)
  =
  \int_{z_-(t^2)}^{z_+(t^2)}\dd z\,
  \overline g_{ia}^{(j)}(t^2,z),
  \qquad
  \overline G_{\mathrm{tot}}^{(j)}
  =
  \sum_{\text{active }i,a}
  \overline G_{ia}^{(j)} .
  \label{eq:overestimate-integrals}
\end{equation}

At each step $j$, the classical side always samples a candidate particle $i$, channel $a$, and energy fraction $z$ from the overestimate distribution.
The composite candidate $\{i,a,z\}$ is drawn from the probability density
\begin{equation}
  p_{\mathrm{prop}}(i,a,z)
  =
  \frac{\overline g_{ia}^{(j)}(t^2,z)}
       {\overline G_{\mathrm{tot}}^{(j)}} .
  \label{eq:proposal-density}
\end{equation}
The quantum veto then corrects this overestimated proposal to the physical density.
The acceptance probability for flavor component $k$ is
\begin{equation}
  P_{\mathrm{acc}}^{(i,a,k)}
  =
  \frac{g_{ia,k}^{(j)}(t^2,z)}
       {p_{\mathrm{prop}}(i,a,z)}
  =
  \frac{g_{ia,k}^{(j)}(t^2,z)\,\overline G_{\mathrm{tot}}^{(j)}}
       {\overline g_{ia}^{(j)}(t^2,z)},
  \label{eq:P-acc}
\end{equation}
and the corresponding rotation angle for the emission qubit is
\begin{equation}
  \theta_{ia,k}^{(j)}
  =
  2\arcsin\sqrt{P_{\mathrm{acc}}^{(i,a,k)}} .
  \label{eq:theta-acc}
\end{equation}
The discretization step size $(-\ln r)$ and the overestimate must be chosen so that $0\leq P_{\mathrm{acc}}^{(i,a,k)}\leq1$ for every populated flavor component.
This is the direct twelve-flavor generalization of the QVPS acceptance condition derived in Ref.~\cite{Bauer:2023ujy}.

For a given proposal $\{i,a,z\}$, the probability that the quantum veto accepts the branching for flavor component $k$ is $P_{\mathrm{acc}}^{(i,a,k)}$, and the probability that it rejects is $1-P_{\mathrm{acc}}^{(i,a,k)}$.
Averaging over all possible proposals, the total emission probability for flavor $k$ at step $j$ is
\begin{align}
  P_{\text{em}}^{(k,j)}
  &=
  \sum_{i,a}\int\!dz\,
  p_{\mathrm{prop}}(i,a,z)\,
  P_{\mathrm{acc}}^{(i,a,k)}
  \nonumber\\
  &=
  \sum_{i,a}\int\!dz\,
  \frac{\overline g_{ia}^{(j)}(t^2,z)}{\overline G_{\mathrm{tot}}^{(j)}}\,
  \frac{g_{ia,k}^{(j)}(t^2,z)\,\overline G_{\mathrm{tot}}^{(j)}}
       {\overline g_{ia}^{(j)}(t^2,z)}
  \;=\;
  \sum_{i,a}\int\!dz\,g_{ia,k}^{(j)}(t^2,z) .
  \label{eq:emission-prob}
\end{align}
The factors of $\overline G_{\mathrm{tot}}^{(j)}$ cancel, and the emission probability reduces exactly to the sum of the physical splitting densities.
Correspondingly, the probability that flavor $k$ survives step $j$ without emitting is
\begin{equation}
  P_{\text{no-em}}^{(k,j)}
  =
  1 - \sum_{i,a}\int\!dz\,g_{ia,k}^{(j)}(t^2,z) .
  \label{eq:no-emission-step}
\end{equation}
Across $J$ steps the survival probability factorizes, and in the limit of fine discretization ($-\ln r\ll 1$) the product exponentiates:
\begin{equation}
  \prod_{j=1}^{J} P_{\text{no-em}}^{(k,j)}
  \;\approx\;
  \exp\!\Bigl[-\sum_{j}\sum_{i,a}\int\!dz\,g_{ia,k}^{(j)}\Bigr]
  \;\longrightarrow\;
  \Delta_k(t_{\mathrm{cut}}^2;t_0^2)~,
  \label{eq:sudakov-from-veto}
\end{equation}
which is the well-known Sudakov form factor. 

The heart of the quantum veto is a controlled rotation that acts on a reusable emission qubit $\ket{e}$, initially in the state $\ket{0}_e$.
If the flavor register is in the state $\sum_k a_k\ket{k}_{\mathrm{int}}$, the gate enacts
\begin{equation}
  \sum_k a_k \ket{k}_{\mathrm{int}}\ket{0}_e
  \;\longrightarrow\;
  \sum_k a_k \ket{k}_{\mathrm{int}}
  \Bigl(
    \cos\frac{\theta_{ia,k}^{(j)}}{2}\ket{0}_e
    +
    \sin\frac{\theta_{ia,k}^{(j)}}{2}\ket{1}_e
  \Bigr).
  \label{eq:quantum-veto-action}
\end{equation}
Measuring the emission qubit yields $\ket{1}_e$ (accept) with probability $\sum_k|a_k|^2\sin^2(\theta_{ia,k}^{(j)}/2)=\sum_k|a_k|^2P_{\mathrm{acc}}^{(i,a,k)}$, which is precisely the correct physical acceptance probability.
Conditioned on the measurement outcome, the flavor amplitudes are multiplied by $\sin(\theta_{ia,k}^{(j)}/2)$ or $\cos(\theta_{ia,k}^{(j)}/2)$, and the classical particle list and kinematics are updated accordingly.
The emission qubit is then reset to $\ket{0}_e$ for the next step.

\subsection{History-space formalism}
We formulate the QVPS process for a fixed number of emissions as an integral over the space of parton-shower histories.
For a given emission combination $(n_g,n_\gamma,n_{\chi})$, where $n_{\chi}$ denotes the number of final produced new $\chi$ pairs, the history set is
\begin{equation}
\Omega(n_g,n_\gamma,n_{\chi})
=
\left\{
h=
\left[
(a_1,z_1,t_1^2),\cdots,(a_{N_h},z_{N_h},t_{N_h}^2)
\right]
\right\},
\label{eq:history-set}
\end{equation}
where $N_h$ is the number of emissions in the history and $a_n$, $z_n$ and $t_n^2$ denote the type, energy fraction and evolution variable of the $n$-th emission, respectively. 
Each history $h$ fixes the observable kinematic information. 

For a fixed history $h$, different flavors generally have different weights for producing that history, denoted by $W_k(h)$.
Conditioned on the history $h$, the probability that the final-state fermion flavor is $k$ satisfies
\begin{equation}
\operatorname{Prob}(k|h)
=
\frac{W_k(h)}{\sum_{k'} W_{k'}(h)}.
\label{eq:prob-k-given-h}
\end{equation}

Since probabilities in a quantum state are given by the squared modulus of amplitudes, we define
\begin{equation}
A_k(h)=\sqrt{W_k(h)}.
\label{eq:amplitude-def}
\end{equation}
The unnormalized final state in flavor space for a fixed history $h$ is then
\begin{equation}
\ket{\Psi(h)}
=
\sum_k A_k(h)\ket{k}_{\mathrm{int}}
\otimes |B(h)\rangle_{\mathrm{pair}}
\label{eq:Psi-h}
\end{equation}
in Hilbert space $\mathcal{V}_{\rm int} \otimes \mathcal{V}_{\rm pair}$, whose squared norm is
\begin{equation}
  \braket{\Psi(h)|\Psi(h)}
  =
  R(h)\sum_k A_k^2(h)
  =
  R(h)\sum_k W_k(h),
  \qquad
  R(h)\equiv \langle B(h)|B(h)\rangle .
  \label{eq:Psi-norm}
\end{equation}
The unobserved pair registers contribute to the history weight a factor \(R(h)\) that is common to all flavors, and the reduced state of the original fermion register remains pure. Its normalized form is
\begin{equation}
\ket{\psi(h)}
=\frac{\sum_{k}A_{k}(h)|k\rangle_{\mathrm{int}}}{\sqrt{\sum_{j}A_{j}^{2}(h)}}.
\label{eq:psi-h}
\end{equation}
To retain the occurrence weight of a history during integration, introduce the unnormalized reduced state
\begin{equation}
|\Phi (h)\rangle = \sum_{k}A_{k}(h)|k\rangle_{\mathrm{int}},\qquad \langle \Phi (h)|\Phi (h)\rangle = \sum_{k}A_{k}^{2}(h) = \sum_{k}W_{k}(h).
\label{eq:Phi-h}
\end{equation}

A central feature of the QVPS framework is the distinction between coherent and incoherent sums.
Different flavor components within the same history $h$ share identical kinematic records and remain coherent, so they are added at the amplitude level.
On the other hand, different histories correspond to distinct emission scales, splitting variables, and orderings. As a result, they are distinguishable kinematic records and must be added incoherently at the density-matrix level.
For fixed $(n_g,n_\gamma,n_\chi)$, the final-state density matrix is therefore
\begin{equation}
\rho(n_g,n_\gamma,{n_\chi})
=
\int_{\Omega(n_g,n_\gamma,{n_\chi})}
dh\,\ket{\Phi(h)}\bra{\Phi(h)}
=
\int_{\Omega(n_g,n_\gamma,{n_\chi})}
dh
\sum_{k,k'}
A_k(h)A_{k'}(h)
\ket{k}\!\bra{k'}_{\mathrm{int}}.
\label{eq:rho-density}
\end{equation}

The total normalization factor is
\begin{equation}
Z_{\mathrm{tot}}
=
\operatorname{Tr}\rho
=
\int_{\Omega(n_g,n_\gamma,{n_\chi})}
dh
\sum_k A_k^2(h).
\label{eq:Z-tot}
\end{equation}
Taking the initial mass-basis state $\ket{\chi_0}$, whose maximal-mixing decomposition is given in Eq.~\eqref{eq:chi0-state}, the weight projected back onto $\ket{\chi_0}$ is
\begin{equation}
Z_{\chi_0}
=
\bra{\chi_0}\rho\ket{\chi_0}
=
\frac{1}{\Nf}
\int_{\Omega(n_g,n_\gamma,{n_\chi})}
dh
\sum_{k,k'}
A_k(h)A_{k'}(h).
\label{eq:Z-chi0}
\end{equation}
For a fixed emission combination, the analytic probability of measuring the initial mass eigenstate $\chi_0$ is
\begin{equation}
P_{\mathrm{Analytic}}(\chi_0|n_g,n_\gamma,n_\chi)
=
\frac{Z_{\chi_0}}{Z_{\mathrm{tot}}}
=
\frac{
\int_{\Omega(n_g,n_\gamma,{n_\chi})}
dh
\left[
\sum_k A_k(h)
\right]^2
}{
\Nf
\int_{\Omega(n_g,n_\gamma,{n_\chi})}
dh
\sum_k A_k^2(h)
}.
\label{eq:P-analytic}
\end{equation}
The expected number of events that fall into the $(n_g,n_\gamma,{n_\chi})$ sector in $N_{\mathrm{shots}}$ independent trials is obtained directly from the total normalization $Z_{\mathrm{tot}}$.
Averaging over the initial flavor distribution yields the event fraction
\begin{equation}
P(n_g,n_\gamma,{n_\chi})
=
\sum_k \frac{1}{\Nf}\int_{\Omega(n_g,n_\gamma,{n_\chi})}dh\,W_k(h)
=
\frac{1}{\Nf}\int_{\Omega(n_g,n_\gamma,{n_\chi})}dh\sum_k W_k(h) .
\label{eq:P-sector}
\end{equation}
Multiplying by the number of trials gives the expected count
\begin{equation}
N(n_g,n_\gamma,{n_\chi})
=
N_{\mathrm{shots}}\,
P(n_g,n_\gamma,{n_\chi})
=
N_{\mathrm{shots}}
\frac{1}{\Nf}
\int_{\Omega(n_g,n_\gamma,{n_\chi})}
dh
\sum_k W_k(h).
\label{eq:N-events}
\end{equation}
The quantity $N(n_g,n_\gamma,n_\chi)$ counts the total event yield in the $(n_g,n_\gamma,n_\chi)$ sector irrespective of the final mass-eigenstate projection.
$P_{\mathrm{Analytic}}(\chi_0|n_g,n_\gamma,n_\chi)$, by contrast, receives its quantum interference signal from the off-diagonal combination $\bigl[\sum_k A_k(h)\bigr]^2 = \sum_{k,k'}A_k A_{k'}$ in its numerator.
The interference between different flavor components, encoded in the $k\neq k'$ cross terms, is the source of the non-trivial mass-eigenstate distribution that constitutes the primary observable of the QVPS algorithm.

\subsection{Computation of history weights and amplitudes}

A parton-shower history consists of two types of factors: the emission vertices that actually occur, and the Sudakov factors describing the absence of additional emissions between adjacent emission scales.
For a given flavor $k$, the differential weight for producing history $h$ factorizes as
\begin{equation}
W_k(h)
=
\prod_{\mathrm{intervals}}\Delta_k
\prod_{\mathrm{emissions}}\omega_k,
\label{eq:W-factorization}
\end{equation}
where $\Delta_k$ denotes the no-emission probability for a given evolution interval, and $\omega_k \equiv \omega_{i\to a, k}(z, t^2)$ are the differential vertex factors defined in Eq.~\eqref{eq:differential_vertices}. 

As a concrete illustration, consider a history with a single dark-photon emission:
\begin{equation}
h=
\left\{
(\chi\to\chi+\gammad,z_1,t_1^2)
\right\}
\in
\Omega(0,1,0).
\label{eq:h-one-gamma}
\end{equation}
This history requires: (i) no splitting during the evolution from the initial scale $t_0^2$ down to $t_1^2$; (ii) one dark-photon emission at $(z_1,t_1^2)$; (iii) the daughter fermion, assigned the reduced starting evolution scale $z_1^2t_1^2$, to evolve to the cutoff without a further emission; and (iv) the dark photon, assigned the starting evolution scale $(1-z_1)^2t_1^2$, to reach the cutoff without conversion into a fermion pair.  The corresponding weight is
\begin{equation}
W_{k}(h) = \Delta_{k}(t_{1}^{2};t_{0}^{2})\,\omega_{\gamma_{D},k}(z_{1},t_{1}^{2})\,\Delta_{k}\big(t_{\mathrm{cut}}^{2};z_{1}^{2}t_{1}^{2}\big)\,\Delta_{\gamma_{D}}\big(t_{\mathrm{cut}}^{2};(1 - z_{1})^{2}t_{1}^{2}\big).
\end{equation}
Note that the last factor is independent of \(k\) and is the elementary example of the factor \(R(h)\) introduced in Eq.~\eqref{eq:Psi-norm}. It reduces to unity when the conversion channel is switched off.

For a history with one gluon and one dark-photon emission, the two time orderings must be computed separately: \(\chi \rightarrow \chi +g\) then \(\chi \rightarrow \chi +\gamma_{\mathrm{D}}\), and \(\chi \rightarrow \chi +\gamma_{\mathrm{D}}\) then \(\chi \rightarrow \chi +g\). For the first ordering, with the gluon emitted at \((z_1,t_1^2)\) and the dark photon at \((z_2,t_2^2)\) where \(t_2^2 < z_1^2 t_1^2\),
\begin{align}
W_{k}(h) ={}& \Delta_{k}(t_{1}^{2};t_{0}^{2})\,\omega_{g}(z_{1},t_{1}^{2})\,\Delta_{k}\big(t_{2}^{2};z_{1}^{2}t_{1}^{2}\big)\,\omega_{\gamma_{D},k}(z_{2},t_{2}^{2})\,\Delta_{k}\big(t_{\mathrm{cut}}^{2};z_{2}^{2}t_{2}^{2}\big) \nonumber\\
& \times \Delta_{g}\big(t_{\mathrm{cut}}^{2};(1 - z_{1})^{2}t_{1}^{2}\big)\,\Delta_{\gamma_{D}}\big(t_{\mathrm{cut}}^{2};(1 - z_{2})^{2}t_{2}^{2}\big).
\end{align}
Here $\Delta_g$ and $\Delta_{\gamma_D}$ denote the no-emission factor of Eq.~\eqref{eq:sudakov} evaluated for the radiated gluon and dark photon, respectively.  The three factors on the first line follow the original fermion line, whose maximum subsequent evolution scale is reset after each emission and which is the only line carrying the flavor label $k$.  The two factors on the second line follow the radiated bosons from their production evolution scales down to the cutoff.  The second ordering is obtained by exchanging the roles of the two vertices. 
The same principle applies to any history: each splitting contributes one vertex factor, and every line contributes Sudakov factors over the intervals between its production, its splittings, and the cutoff.

\subsection{Qubit encoding and mass-basis readout}
Each fermion flavor is stored in four qubits.
Computational states $\ket{0},\ldots,\ket{11}$ represent the twelve physical interaction eigenstates, while $\ket{12},\ldots,\ket{15}$ are inactive and act as the identity under all shower operations.
The $12\times12$ mass mixing matrix $U$ in Eq.~\eqref{eq:parameterized-mixing} is embedded into a $16\times16$ unitary:
\begin{equation}
  \widetilde V
  =
  U\oplus I_4 .
  \label{eq:embedded-dft}
\end{equation}

\begin{figure}[H]
\centering
\begin{quantikz}[row sep={0.42cm,between origins}, column sep=0.55cm]
\lstick{$\ket{f_3}$} & \gate[wires=4]{\widetilde V} & \meter{} \\
\lstick{$\ket{f_2}$} &                              & \meter{} \\
\lstick{$\ket{f_1}$} &                              & \meter{} \\
\lstick{$\ket{f_0}$} &                              & \meter{}
\end{quantikz}
\caption{Mass-basis readout of one four-qubit fermion register at the end of the event.}
\label{fig:dft-basis}
\end{figure}
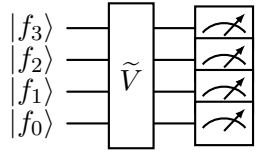

The shower is evolved in the interaction basis because all charge-dependent splittings are diagonal there.
At the end of the event, $\widetilde V$ is applied to the observed fermion register and the four qubits are measured, as shown in Fig.~\ref{fig:dft-basis}.

\subsection{Quantum circuits for the veto algorithm}
\begin{figure}[H]
\centering
\begin{quantikz}[row sep={0.45cm,between origins}, column sep=0.5cm]
\lstick{$\ket{f_3}$} & \gate[wires=5]{U_a^{(j)}(t_j^2,z)} & \qw \\
\lstick{$\ket{f_2}$} &                                    & \qw \\
\lstick{$\ket{f_1}$} &                                    & \qw \\
\lstick{$\ket{f_0}$} &                                    & \qw \\
\lstick{$\ket{0}_e$} &                                    & \meter{} & \rstick{$b_j$}
\end{quantikz}
\caption{One QVPS veto step after classical sampling of the candidate particle, channel, evolution scale, and energy fraction.  The five-wire gate is either flavor-independent or flavor-conditioned.  The emission qubit is measured and reset before the next step.}
\label{fig:qvps-step}
\end{figure}
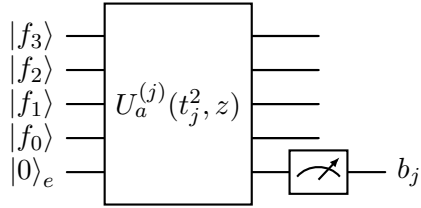

Once the classical registers have fixed $(i,a,t_j^2,z)$, only the flavor register $f_i$ of the candidate fermion and the reusable emission qubit $e$ are needed for the veto.
Figure~\ref{fig:qvps-step} shows the register-level circuit.
Measurement outcome $b_j=1$ accepts the splitting; $b_j=0$ rejects it.
The result and sampled kinematics are saved in the classical history, after which $e$ is reset for the next step.

For the channels with a flavor-independent veto angle
$\chi\to\chi g$,
$g\to gg$,
$g\to\chi\bar{\chi}$,
$\gammad\to\chi\bar{\chi}$,
the gate reduces to a single rotation on the emission qubit, applied uniformly to all flavor components:
\begin{equation}
  U_a^{(j)}
  =
  I_F\otimes R_Y(\theta_{ia,k}^{(j)}), 
  \label{eq:flavor-independent-veto}
\end{equation}
with the $\theta_{ia,k}^{(j)}$ defined in Eq.~\eqref{eq:theta-acc}.
The flavor amplitudes are unchanged by the accept/reject decision.

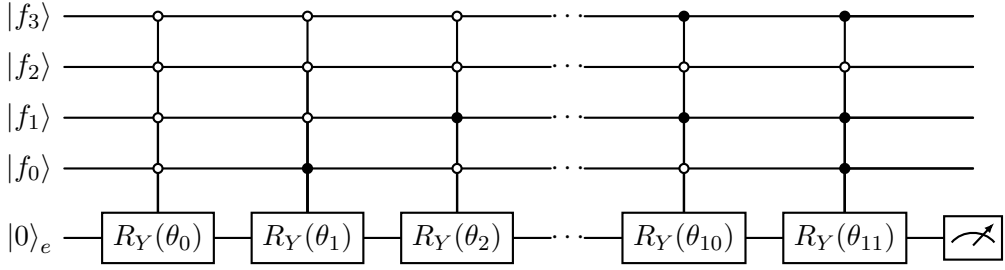
\begin{figure}[htb]
\centering
\begin{quantikz}
\lstick{$\ket{f_3}$} &\octrl{4} & \octrl{4} & \octrl{4} & \push{\cdots} & \ctrl{4}  & \ctrl{4}  & \qw \\
\lstick{$\ket{f_2}$}
& \octrl{3} & \octrl{3} & \octrl{3} & \push{\cdots} & \octrl{3} & \octrl{3} & \qw \\
\lstick{$\ket{f_1}$}
& \octrl{2} & \octrl{2} & \ctrl{2}  & \push{\cdots} & \ctrl{2}  & \ctrl{2}  & \qw \\
\lstick{$\ket{f_0}$}
& \octrl{1} & \ctrl{1}  & \octrl{1} & \push{\cdots} & \octrl{1} & \ctrl{1}  & \qw \\
\lstick{$\ket{0}_e$}
& \gate{R_Y(\theta_0)}
& \gate{R_Y(\theta_1)}
& \gate{R_Y(\theta_2)}
& \push{\cdots}
& \gate{R_Y(\theta_{10})}
& \gate{R_Y(\theta_{11})}
& \meter{}
\end{quantikz}%
\caption{Multiplexed veto for $\chi\to\chi\gammad$.  Each physical interaction eigenstate controls its own rotation of the emission qubit. Here, the $\theta_k$ denotes $\theta_{\gammad,k}^{(j)}$ for $k=0,\cdots,11$. 
\label{fig:flavor-dependent-veto}}
\end{figure}

For the flavor-dependent channel $\chi\to\chi\gammad$, Eq.~\eqref{eq:theta-acc} gives a different angle for every interaction eigenstate.
The corresponding unitary is block-diagonal in the flavor basis:
\begin{equation}
  U_{\gammad}^{(j)}
  =
  \sum_{k=0}^{11}
  \ket{k}\!\bra{k}
  \otimes R_Y(\theta_{\gammad,k}^{(j)})
  +
  \sum_{k=12}^{15}
  \ket{k}\!\bra{k}
  \otimes I_2 .
  \label{eq:block-veto}
\end{equation}
This five-qubit $32\times32$ gate is implemented in the statevector simulator as a direct matrix operation.
Its decomposition into controlled rotations is displayed in Fig.~\ref{fig:flavor-dependent-veto}, in which open and filled controls select the binary representation of the interaction eigenstate index $k$.

Conditioning on the measurement outcome multiplies the amplitude of flavor $k$ by either
$\sin(\theta_{\gammad,k}^{(j)}/2)$
or
$\cos(\theta_{\gammad,k}^{(j)}/2)$.
Repeated dark-photon vetoes therefore change the relative flavor amplitudes, which the final application of $\widetilde V$ converts into the mass-eigenstate interference pattern studied in Sec.~\ref{sec:results}.

\subsection{Pair production and event update}
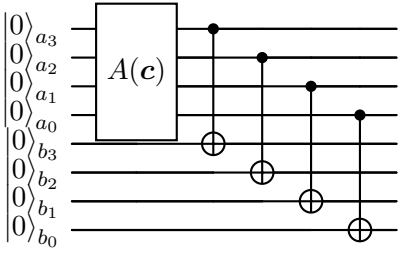
\begin{figure}[htb]
\centering
\begin{quantikz}[row sep={0.38cm,between origins}, column sep=0.32cm]
\lstick{$\ket{0}_{a_3}$} & \gate[wires=4]{A(\boldsymbol c)} & \ctrl{4} & \qw      & \qw      & \qw      & \qw \\
\lstick{$\ket{0}_{a_2}$} &                                  & \qw      & \ctrl{4} & \qw      & \qw      & \qw \\
\lstick{$\ket{0}_{a_1}$} &                                  & \qw      & \qw      & \ctrl{4} & \qw      & \qw \\
\lstick{$\ket{0}_{a_0}$} &                                  & \qw      & \qw      & \qw      & \ctrl{4} & \qw \\
\lstick{$\ket{0}_{b_3}$} & \qw                              & \targ{}  & \qw      & \qw      & \qw      & \qw \\
\lstick{$\ket{0}_{b_2}$} & \qw                              & \qw      & \targ{}  & \qw      & \qw      & \qw \\
\lstick{$\ket{0}_{b_1}$} & \qw                              & \qw      & \qw      & \targ{}  & \qw      & \qw \\
\lstick{$\ket{0}_{b_0}$} & \qw                              & \qw      & \qw      & \qw      & \targ{}  & \qw
\end{quantikz}
\caption{Preparation of the correlated flavor state in Eq.~\eqref{eq:pair-state}.  The amplitude-preparation gate does not affect computational states $12,\ldots,15$.
\label{fig:pair-preparation}}
\end{figure}

If a boson-to-fermion-pair proposal is accepted, two new four-qubit registers are appended in the entangled state
\begin{equation}
  \ket{\Psi_{\chi\bar{\chi}}}
  =
  \sum_{k=0}^{11}c_k\,
  \ket{k}_{\chi}\ket{k}_{\bar{\chi}},
  \label{eq:pair-state}
\end{equation}
with flavor coefficients determined by the production channel:
\begin{align}
  c_k^{(g)}
  =
  \frac{1}{\sqrt{12}}
  \qquad  {\rm for }~ g \to \chi \bar{\chi} ~,  \nonumber\\
  c_k^{(\gammad)}
  =
  \frac{\sqrt{\alphad Q_k^2}}
       {\sqrt{\sum_{\ell=0}^{11}\alphad Q_\ell^2}}    \qquad  {\rm for }~ \gamma_D \to \chi \bar{\chi}  ~. \label{eq:pair-coefficients}
\end{align}
The circuit in Fig.~\ref{fig:pair-preparation} operates in two stages.
First, the amplitude-preparation gate $A(\boldsymbol c)$ prepares the coherent superposition $\sum_k c_k\ket{k}_a$ in register $a$ from the initial $\ket{0}$ state, with coefficients given by Eq.~\eqref{eq:pair-coefficients}.
Four CNOT gates then correlate the two registers: register $a$ acts as the control and register $b$ as the target, one qubit pair per bit of the four-qubit flavor label.
Under this operation each computational basis state transforms as $\ket{k}_a\ket{0}_b\to\ket{k}_a\ket{k}_b$, producing the entangled pair state $\sum_k c_k\ket{k}_a\ket{k}_b$.

In the quantum simulation, each accepted vector to fermion splitting would append two four-qubit flavor registers to the statevector, increasing its dimension and memory footprint by a factor of $2^8=256$ for every produced pair on a classical computer.
Repeated pair production therefore rapidly exceeds the practical memory budget of the present statevector simulator.
Although our simulation framework implements the Equations~\eqref{eq:pair-state} and \eqref{eq:pair-coefficients}, together with Fig.~\ref{fig:pair-preparation}, the numerical results reported in the next section do not include either $g\to\chi\bar{\chi}$ or $\gammad\to\chi\bar{\chi}$ for the practical issue. 

To summarize, the complete event update integrates the components introduced in the preceding subsections into a single hybrid loop.
At each evolution step $j$ the classical routine lowers the active-particle scales, samples a particle $i$, a channel $a$, and an energy fraction $z$ from the overestimate distribution of Eq.~\eqref{eq:proposal-density}, and computes the acceptance angles of Eq.~\eqref{eq:theta-acc}.
These values are supplied as classical control parameters to the appropriate gate in Fig.~\ref{fig:qvps-step}.
The quantum veto then takes over: the controlled rotation of Eq.~\eqref{eq:quantum-veto-action} entangles the flavor register with the emission qubit, and the measurement of the emission qubit implements the flavor-dependent accept/reject decision.
Conditioned on the outcome, each flavor component $k$ is multiplied by $\sin(\theta_{ia,k}^{(j)}/2)$ (accept) or $\cos(\theta_{ia,k}^{(j)}/2)$ (reject).
Channels with a flavor-independent veto angle ($\chi\to\chi g$, $g\to gg$, $g\to\chi\bar\chi$ and $\gammad\to\chi\bar\chi$) preserve the relative amplitudes across different flavors, whereas the flavor-dependent $\chi\to\chi\gammad$ channel applies a $Q_k$-dependent angle that reweights the flavor composition differently for each $k$. This effect has no classical analogue.
The cumulative reweighting over many veto steps builds up non-uniform flavor amplitudes, and the final maximal-mixing basis rotation into the mass basis converts this charge-dependent modulation into an interference pattern among the mass eigenstates. This is the quantum signal that $P_{\mathrm{Analytic}}(\chi_0|n_g,n_\gamma)$ is designed to capture, and the analytic verification below tests it directly.
After the measurement, the classical particle list and kinematics are updated accordingly.
The classical-proposal--quantum-veto loop repeats until the evolution reaches $t^2_{\mathrm{cut}}=4\Mq^2$, after which the observed fermion register is rotated into the mass basis and measured.

\subsection{Analytic verification in simple cases}
To verify that the quantum circuits correctly implement the history-space formalism, we perform an analytic comparison in a simplified early-splitting scenario before showing the results of comprehensive simulation.
The uniform coherent pure initial state as defined in Eq.~(\ref{eq:chi0-state}) is adopted. 
The evolution is restricted to the range $1~\text{GeV}^2\le t^2\le t_0^2=100~\text{GeV}^2$, corresponding to the first a few steps of the shower.
In the analytic calculation, all emission-scale integrals and Sudakov factors are evaluated in this range. The quantum simulation results are collected over the same scale window.

We choose the following parameters for illustration:
\begin{equation}
\Nf=12,
\qquad
\alphad=0.2,
\qquad
C_F=\frac{4}{3},
\qquad
\Lambda=0.2~\text{GeV},
\qquad
N_{\mathrm{shots}}=100000,
\label{eq:verification-params}
\end{equation}
the analytic decohered form [Eq.~\eqref{eq:P-analytic}] and the statistical results from the QVPS quantum simulation are compared in Table~\ref{tab:verification}.

\begin{table}[h]
\centering
\begin{tabular}{|c|cc|cc|} \hline
$(n_g,n_\gamma)$
& $P_{\mathrm{Analytic}}$
& $P_{\mathrm{QVPS}}$
& $N_{\mathrm{Analytic}}$
& $N_{\mathrm{QVPS}}$ \\ \hline
$(0,0)$ & 0.6327 & 0.7308 & 86.71 & 78 \\
$(0,1)$ & 0.9206 & 0.9176 & 659.48 & 643 \\
$(1,0)$ & 0.6496 & 0.6482 & 525.71 & 523 \\
$(1,1)$ & 0.9056 & 0.8891 & 1248.41 & 1244 \\ \hline
\end{tabular}
\caption{Comparison of analytic and QVPS results for the early-splitting stage ($1\le t^2\le 100$).  $P$ (with $P_{\rm Analytic}$ defined in Eq.~\ref{eq:P-analytic}) denotes the probability of measuring the initial mass eigenstate $\chi_0$, and $N$ (with $N_{\rm Analytic}$ defined in Eq.~\ref{eq:N-events}) is the expected number of events in $N_{\mathrm{shots}}=100000$ trials. \label{tab:verification}}
\end{table}

The agreement between the analytic decohered form and the QVPS simulation is excellent for all channels except $(0,0)$, where the small number of expected events ($\sim87$) leads to a larger relative statistical fluctuation.
This comparison validates two essential features of the QVPS construction.
First, the flavor coherence within each history is correctly preserved by the quantum circuits. The interference term $\sum_{k,k'}A_kA_{k'}$ in Eq.~\eqref{eq:P-analytic} is essential for the agreement, while a purely classical treatment would give a different result.
Second, the different kinematic histories are indeed added at the density-matrix level rather than the amplitude level, so that the history decoherence structure is consistent with the statistical ensemble produced by the repeated quantum measurements.

\section{Numerical Results}
\label{sec:results}

We now present parton-level results for the finite-mass conformal-window shower.
Unless stated otherwise, the classical and QVPS samples use the same splitting
kernels, running couplings, phase-space limits, and evolution grid.  The enabled
channels are $\chi\to\chi g$, $\chi\to\chi\gammad$, and $g\to gg$. The
vector-to-fermion channels are omitted because of the statevector-memory
limitation discussed in Sec.~\ref{sec:qvps}.  
Throughout the work, we take the initial $\chi$ to have energy $E=50$ GeV (traveling along the Z axis), $\Lambda=0.2$ GeV, and $t_{\max}^2=100$ GeV$^2$.
The gluon and dark photon are treated as massless, except for an infrared evolution-scale cutoff introduced to regularize the soft divergence.
The comparison therefore isolates
the effect of replacing the classical flavor update by the charge-dependent
quantum veto and the final coherent mass-basis readout within the present
framework.

\subsection{Energy-Energy correlations and angular scaling}
To investigate the angular scaling behavior of the simulated dark shower, we measure the all-pairs, absolute energy-energy correlation (EEC)~\cite{Basham:1978zq,Basham:1978bw}:
\begin{equation}
\frac{\mathrm{d}\Sigma_{\text{abs}}}{\mathrm{d}R_L} = \frac{1}{N_{\text{evt}}} \sum_{e=1}^{N_{\text{evt}}} \sum_{i < j \in e} 2E_i E_j \, \delta(R_L - \Delta R_{ij}),
\label{eq:eec_def}
\end{equation}
where $E_i$ is the energy of particle $i$, and $\Delta R_{ij} \approx \theta_{ij}$ is the opening angle between particles $i$ and $j$ in the collinear limit.

As discussed in Sec.~\ref{sec:framework}, the dark $\SU(3)_D$ coupling approaches the fixed point of Eq.~\eqref{eq:alpha-star} in the infrared regime. 
When the coupling is approximately frozen ($\beta(\alpha_s) \approx 0$) and the dark quarks can be treated as effectively massless, the theory has no intrinsic dimensional scale. Consequently, collinear observables will transform as power laws under angular rescaling~\cite{Hofman:2008ar,Dixon:2019uzg,Korchemsky:2019nzm}:
\begin{equation}
F(R_L) \equiv \frac{\mathrm{d}\Sigma_{\text{abs}}}{\mathrm{d}R_L} \propto R_L^{\eta_*},
\label{eq:scaling}
\end{equation}
where the $\eta_*$ encodes both the canonical dimension of collinear phase space, and the anomalous dimension generated by multiple emissions. On a double-logarithmic plot, Eq.~\eqref{eq:scaling} gives a straight line with slope $\eta_*$:
\begin{equation}
\ln F(R_L) = \eta_* \times \ln R_L + \text{const}.
\end{equation}

\begin{figure}[htb]
  \centering
  \includegraphics[width=0.6\linewidth]{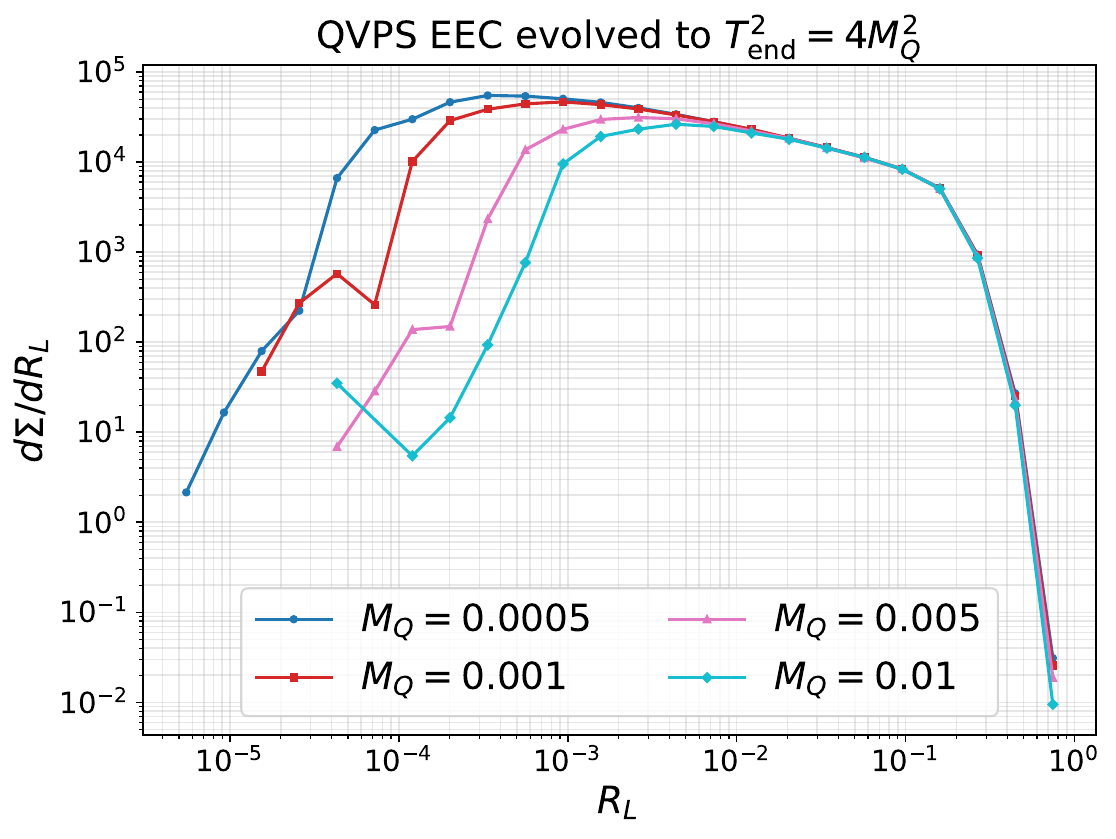}
  \caption{All-pairs absolute EEC for QVPS samples with $E_0 = 50$ GeV, $\Lambda_D = 0.2$ GeV, and four choices of dark-quark mass $M_q \in [5\times 10^{-4}~\text{GeV}, 10^{-2}~\text{GeV}]$.
  \label{fig:result-eec}}
\end{figure}

Figure~\ref{fig:result-eec} displays the measured EEC for several values of the dark-quark mass $M_q$ based on our quantum simulation. 
The approximately linear trend over the intermediate angular range ($10^{-3} \lesssim R_L \lesssim 10^{-1}$) in double-logarithmic coordinates suggests approximate power-law scaling over this angular interval.
Such behavior is qualitatively compatible with scaling in a near-conformal regime. Outside this intermediate window, departures from the power-law scaling are associated with mass-dependent phase-space restrictions and wide-angle kinematic boundaries. 
For small angles ($R_L \lesssim M_q / E_0$), the dark quarks carry a non-zero mass $M_q$, which terminates the shower evolution at the kinematic cutoff $t_{\mathrm{cut}}^2 = 4M_q^2$. 
This limits the angular range available for the power-law scaling and produces the turnover seen on the left side of Fig.~\ref{fig:result-eec}. As expected, lighter quarks allow the shower to evolve further, pushing the turnover toward smaller angles, whereas larger quark masses shift the turnover toward larger $R_L$.
At wide angles ($R_L \to 1$), the collinear approximation ($\theta \ll 1$) breaks down, and total energy-momentum conservation restricts wide-angle radiation. This produces the steep falloff at the right boundary.

\subsection{Classical versus Quantum}
\label{sec:classical-vs-quantum}
We construct a classical parton shower algorithm that shares the identical kinematic conditions with the QVPS, to highlight the observable signatures of quantum interference. 
The distinction lies exclusively in the treatment of internal flavor degrees of freedom. For the classical Markovian evolution, the radiating parton propagates a definite physical flavor label in the mass basis. Upon an accepted dark-photon emission $\chi_i \to \chi_j \gammad$, the daughter flavor $j$ is sampled from the classical transition probability
  \begin{equation}
    P_{ji} = \frac{|\widehat{Q}_{ji}|^2}{\sum_{k} |\widehat{Q}_{ki}|^2}, 
    \qquad 
    \widehat{Q}(\xi) = U(\xi)\,\mathrm{diag}(Q_0, \dots, Q_{11})\,U^\dagger(\xi),
    \label{eq:rotated-charge-matrix}
  \end{equation}
where $U(\xi)$ is the flavor mixing matrix.
This projective sampling destroys all quantum relative phases between flavor components at each branching vertex.
The numerical results presented below therefore provide a controlled comparison of coherent quantum flavor interference against classical Markovian evolution.

\begin{figure}[htb]
  \centering
  \includegraphics[width=0.6\linewidth]{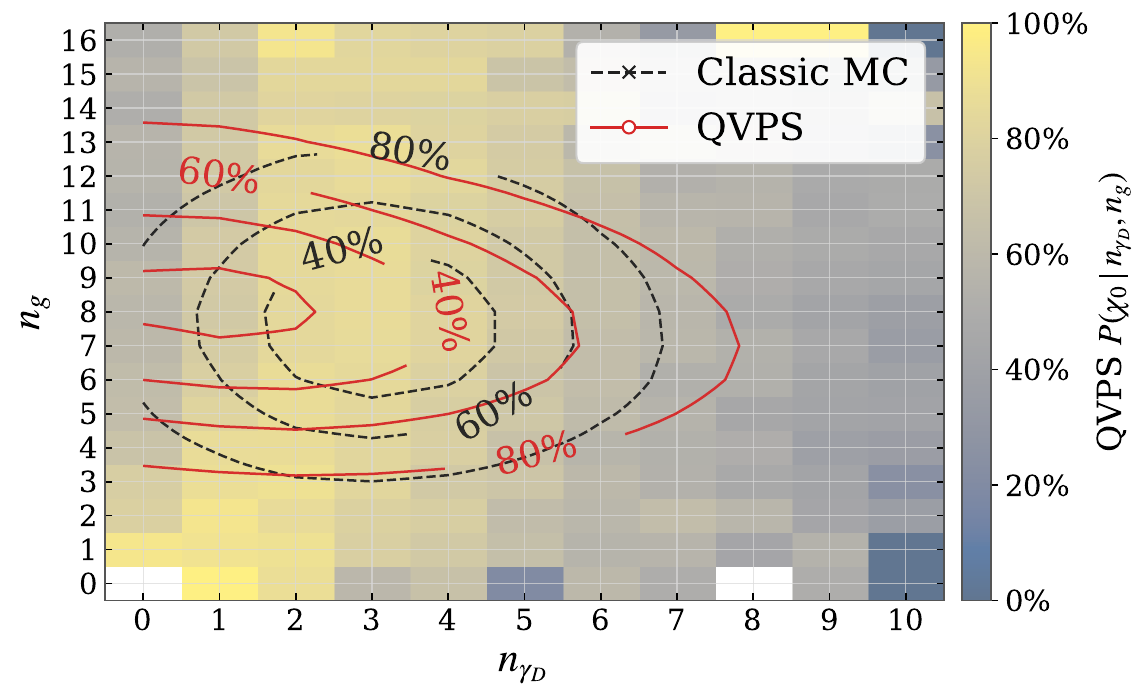}
  \caption{Joint distribution of the dark-photon and dark-gluon multiplicities. Dashed black (classical MC) and solid red (QVPS) contours enclose the highest-density 20\%, 40\%, 60\%, and 80\% regions. The displayed sample means are $\langle n_{\gammad}\rangle = 3.344$ (Classical) versus $3.216$ (QVPS), and $\langle n_g\rangle = 7.869$ (Classical) versus $7.988$ (QVPS).
  \label{fig:result-emission-counts}}
\end{figure}

The global characteristics of the shower activity are summarized by the joint multiplicity contours in Fig.~\ref{fig:result-emission-counts}. Both algorithms exhibit substantial geometric overlap across the core emission region, confirming that the overall shower intensity is preserved. While the dark-gluon mean is similar between the Classical and QVPS samples ($\langle n_g \rangle = 7.869 \to 7.988$), the dark-photon multiplicity displays a noticeably larger reduction ($\langle n_{\gammad} \rangle = 3.344 \to 3.216$). 

\begin{figure}[htb]
  \centering
    \includegraphics[width=0.4\linewidth]{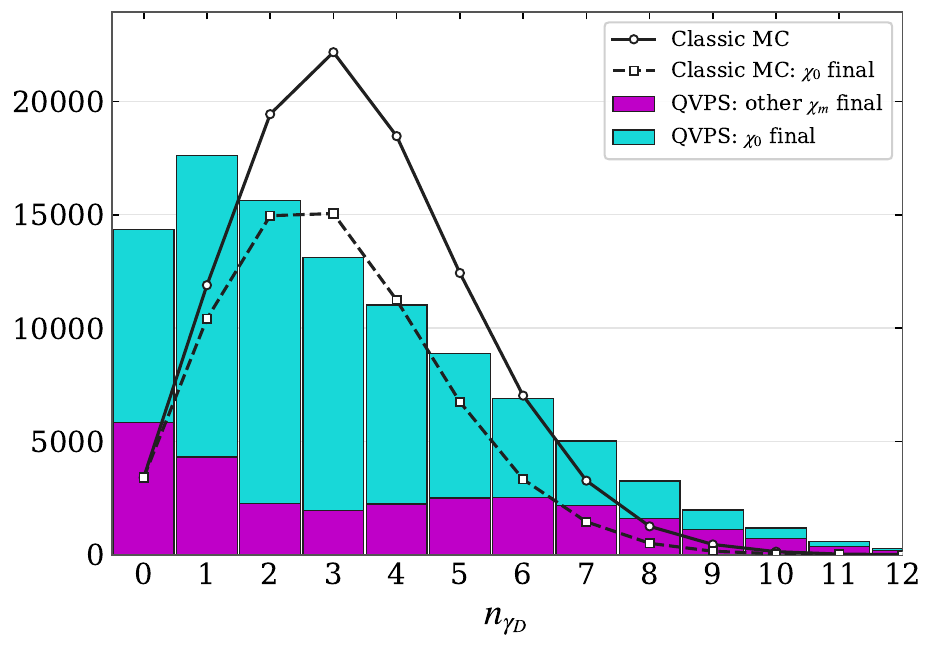}
    \includegraphics[width=0.4\linewidth]{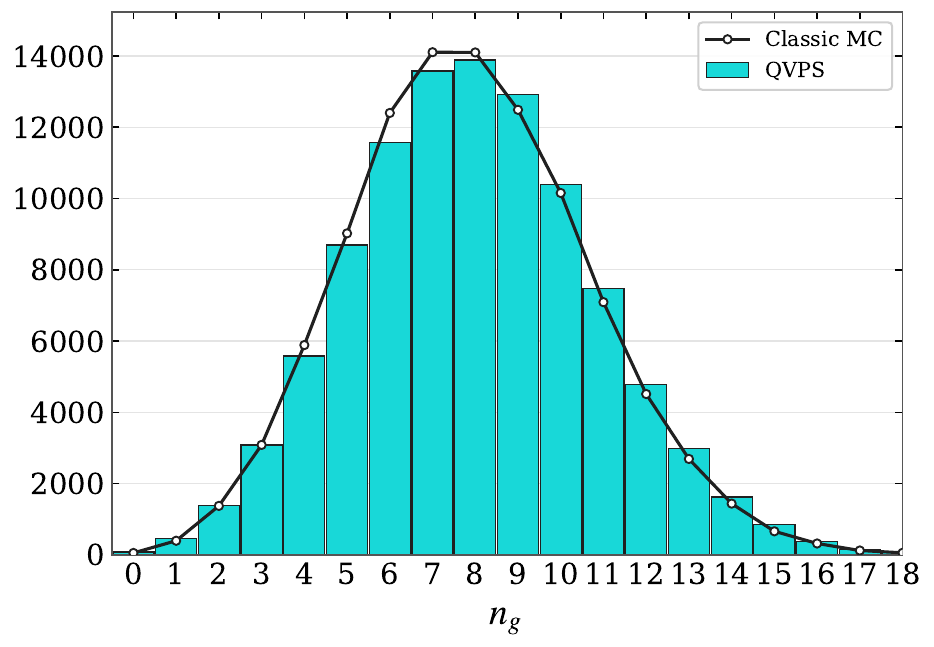}
  \caption{One-dimensional projections of the multiplicity distributions. Left panel: Dark-photon multiplicity $n_{\gammad}$, where the QVPS/classical sample is further separated into events terminating in the ground mass state $\chi_0$ versus other states $\chi_{m, ~m \neq 0}$. Right: Dark-gluon multiplicity $n_g$ comparing inclusive classical Monte Carlo and QVPS distributions.
  \label{fig:result-multiplicity-projections}}
\end{figure}

The physical origin of this behavior can be more easily seen in the one-dimensional projections in Fig.~\ref{fig:result-multiplicity-projections}.
In the right panel of Fig.~\ref{fig:result-multiplicity-projections}, both simulations yield nearly identical dark-gluon distributions peaking around $n_g = 7$--$8$. In our model, $SU(3)_D$ color interactions are strictly flavor-blind. Consequently, the rotation angle $\theta_{ia,k}^{(j)}$ applied to the emission qubit is independent of the flavor index $k$. The veto step acts as an overall scalar factor on the flavor state, leaving the relative flavor amplitudes and quantum phases unperturbed. A small indirect effect is nevertheless expected from the emission of dark photons, which changes the emitter's subsequent evolution scale and energy fraction and thereby shifts the kinematic phase space available for subsequent gluon radiation. 

In contrast, the dark-photon emissions ($\chi \to \chi \gammad$) depend quadratically on the charge spectrum $Q_k \in [0.5, 2.0]$. In QVPS, the accept and reject outcomes dynamically reweight the statevector:
  \begin{equation}
    |k\rangle_{\mathrm{int}} \;\xrightarrow{\text{accept}}\; \sin\!\left(\frac{\theta_{\gammad, k}^{(j)}}{2}\right) |k\rangle_{\mathrm{int}},
    \qquad
    |k\rangle_{\mathrm{int}} \;\xrightarrow{\text{reject}}\; \cos\!\left(\frac{\theta_{\gammad, k}^{(j)}}{2}\right) |k\rangle_{\mathrm{int}}.
  \end{equation}
  Because $\sin(\theta_{\gammad, k}^{(j)}/2) \propto Q_k$, observing no emission enhances the relative amplitude of smaller-$Q_k$ components. If a shower history experiences no emissions early on, the fermion's effective charge is dynamically reduced, suppressing the emission probability in subsequent evolution steps. This coherent feedback mechanism produces a significantly higher fraction of zero-emission events in QVPS ($14.54\%$ versus $3.38\%$ classically) and shifts the peak from $n_{\gammad} = 3$ down to $n_{\gammad} = 1$. Conversely, once an emission is accepted, the amplitude shifts toward higher-$Q_k$ components, enhancing the emission rate for later steps and yielding a broader distribution with an extended high-multiplicity tail.
 
Summing over all multiplicity bins in the left panel of Fig.~\ref{fig:result-multiplicity-projections}, the probability of finding the emitter in the ground mass eigenstate $\chi_0$ at the end of the shower is $72.17\%$ in QVPS compared to $67.10\%$ in the classical Markov shower, an absolute increase of $\sim5$ percentage points. In QVPS the initial state is the uniform coherent superposition of Eq.~\eqref{eq:chi0-state}. Interference among different flavor amplitudes preserves a larger return probability to the initial state than does an incoherent classical sum over independent Markov transition probabilities $|\widehat{Q}_{ji}|^2$.

\begin{figure}[htb]
  \centering
  \includegraphics[width=0.68\linewidth]{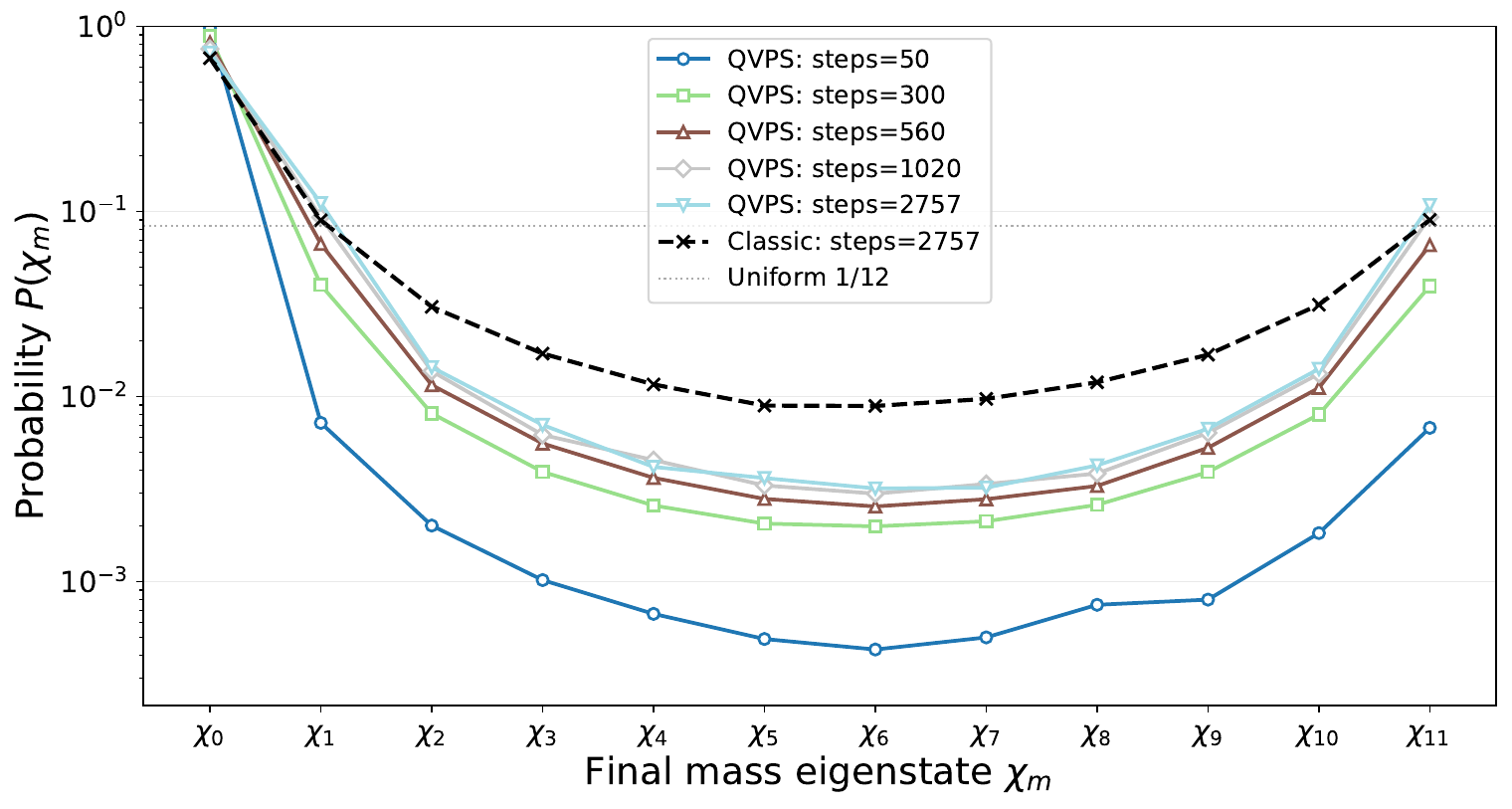}
  \caption{Twelve-state mass eigenstate distribution for classical Monte Carlo (black dashed curve, final step) and QVPS (across intermediate and final evolution stages) at maximal mixing ($\xi=1$). The horizontal dashed line at  1/12 represents the uniform incoherent reference baseline.
  \label{fig:mass_state_compare}}
\end{figure}

The clearest manifestation of coherent flavor evolution appears in the final mass-eigenstate distribution $p(\chi_m)$, shown in Fig.~\ref{fig:mass_state_compare}. Both models exhibit a primary peak at the initial state $\chi_0$ and decay toward the central states. Comparing to the classical case, the QVPS substantially suppresses intermediate modes ($m \in [3, 8]$) while amplifying the boundary states $\chi_0$ and $\chi_{11}$.
To understand the origin of this characteristic U-shaped distribution and its deviation from classical Markovian evolution, we derive the post-shower mass spectrum analytically using the density-matrix formalism. 
 
Consider an arbitrary density matrix in the interaction basis, $\rho = \sum_{k,k'} \rho_{kk'} |k\rangle\langle k'|$. At step $j$, the controlled veto unitary entangles the candidate fermion with the emission qubit:
  \begin{equation}
    |k\rangle_{\mathrm{int}} |0\rangle_e 
    \;\longrightarrow\; 
    \sqrt{1 - p_{j,k}}\,|k\rangle_{\mathrm{int}}|0\rangle_e 
    + 
    \sqrt{p_{j,k}}\,|k\rangle_{\mathrm{int}}|1\rangle_e,
  \end{equation}
  where $p_{j,k} \equiv \sin^2(\theta_{\gammad, k}^{(j)}/2)$ is the physical branching probability for component $k$. Projective measurement of the emission qubit yields outcome $b_j = 1$ (accept) or $b_j = 0$ (reject), described by the measurement Kraus operators:
  \begin{equation}
    M_1 = \sum_{k} \sqrt{p_{j,k}}\,|k\rangle\langle k|,
    \qquad
    M_0 = \sum_{k} \sqrt{1 - p_{j,k}}\,|k\rangle\langle k|.
  \end{equation}
  Summing over the unobserved emission outcomes, the ensemble-averaged density matrix evolves as $\rho' = M_0 \rho M_0^\dagger + M_1 \rho M_1^\dagger$. In matrix components,
  \begin{equation}
    \rho'_{kk'} 
    = 
    \langle k | \rho' | k' \rangle 
    = 
    \eta_{j,kk'}\,\rho_{kk'},
    \label{eq:step-decoherence}
  \end{equation}
  where the single-step coherence factor is
  \begin{equation}
    \eta_{j,kk'} = \sqrt{p_{j,k} p_{j,k'}} + \sqrt{(1 - p_{j,k})(1 - p_{j,k'})}.
    \label{eq:mass-state-coherence-factor}
  \end{equation}

In a fine shower discretization ($-\ln r \ll 1$), the branching probability per step is small, $p_{j,k} \simeq \lambda_j Q_k^2 \ll 1$, where $\lambda_j$ encapsulates the phase-space volume and coupling factors at step $j$. Expanding Eq.~\eqref{eq:mass-state-coherence-factor} to first order in $\lambda_j$:
  \begin{align}
    \sqrt{p_{j,k} p_{j,k'}} 
    &= \lambda_j Q_k Q_{k'}, \nonumber \\
    \sqrt{(1 - p_{j,k})(1 - p_{j,k'})} 
    &\simeq 
    1 - \frac{1}{2}(p_{j,k} + p_{j,k'}) 
    = 1 - \frac{1}{2}\lambda_j (Q_k^2 + Q_{k'}^2).
  \end{align}
  Combining these two contributions yields
  \begin{equation}
    \eta_{j,kk'} 
    \simeq 
    1 - \frac{1}{2}\lambda_j \left(Q_k^2 - 2Q_k Q_{k'} + Q_{k'}^2\right)
    = 
    1 - \frac{1}{2}\lambda_j (Q_k - Q_{k'})^2.
  \end{equation}
  Using the uniformly spaced benchmark charges $Q_k = Q_{\min} + k\,\delta Q$, we have $Q_k - Q_{k'} = (k - k')\,\delta Q$. Exponentiating across $J$ successive evolution intervals, the accumulated off-diagonal suppression takes a Gaussian form:
  \begin{equation}
    \rho_{kk'}^{(J)} 
    = 
    \rho_{kk'}^{(0)} \prod_{j=1}^J \eta_{j,kk'} 
    \simeq 
    \frac{1}{12}\exp\!\left[ -a_J (k - k')^2 \right],
    \qquad
    a_J \equiv \frac{(\delta Q)^2}{2}\sum_{j=1}^J \lambda_j,
    \label{eq:mass-state-coherence-envelope}
  \end{equation}
  where we evaluated the initial uniform density matrix $\rho_{kk'}^{(0)} = 1/12$ corresponding to $|\chi_0\rangle = \frac{1}{\sqrt{12}}\sum_k |k\rangle_{\mathrm{int}}$.

At the end of the shower, the state is projected onto mass eigenstate $|m\rangle_{\mathrm{mass}} = \frac{1}{\sqrt{12}}\sum_{k=0}^{11} e^{-i 2\pi m k / 12} |k\rangle_{\mathrm{int}}$. The probability of measuring state $\chi_m$ is
  \begin{equation}
    P_m(J) 
    = 
    \langle m | \rho^{(J)} | m \rangle 
    = 
    \frac{1}{12}\sum_{k=0}^{11} \sum_{k'=0}^{11} e^{i \frac{2\pi m (k - k')}{12}} \rho_{kk'}^{(J)}.
  \end{equation}
  Substituting Eq.~\eqref{eq:mass-state-coherence-envelope} and defining the index distance $d = |k - k'|$:
  \begin{itemize}
    \item For $d = 0$ ($k = k'$), there are $12$ diagonal entries, contributing $12 \times 1 = 12$.
    \item For each distance $d \in \{1, \dots, 11\}$, there are $12 - d$ pairs with $k - k' = d$ and $12 - d$ pairs with $k - k' = -d$. Combining the phase factors gives $e^{-i 2\pi m d / 12} + e^{i 2\pi m d / 12} = 2\cos(2\pi m d / 12)$.
  \end{itemize}
  Grouping terms by separation distance $d$ yields the closed-form mass spectrum:
  \begin{equation}
    P_m(J)
    \simeq
    \frac{1}{12^2}
    \left[
      12
      + 2\sum_{d=1}^{11}(12 - d)\,e^{-a_J d^2}
        \cos\!\left(\frac{2\pi m d}{12}\right)
    \right].
    \label{eq:mass-state-asymptotic-spectrum}
  \end{equation}

Equation~\eqref{eq:mass-state-asymptotic-spectrum} provides transparent insights into the physical limiting behaviors. 
In a physical shower, evolution terminates at the infrared cutoff scale $t_{\mathrm{cut}}^2 = 4M_q^2$, freezing $a_J$ at a finite value. Because the mass index $m$ is a discrete Fourier mode defined modulo 12, mode $m = 11 \equiv -1 \pmod{12}$ is topologically adjacent to $m = 0$ on the unit circle. Reflection symmetry of the cosine implies $P_m = P_{12-m}$. Plotting the spectrum on a linear axis $m \in [0, 11]$ splits the periodic peak centered at $m = 0$, placing its symmetric shoulders at $m = 0$ and $m = 11$. The central states near $m = 6$ lie farthest from the origin on the Fourier circle and experience maximal destructive interference from the alternating sign of $\cos(\pi d) = (-1)^d$, creating the trough of the U-shaped curve.

In the fully coherent limit ($a_J \to 0$), the sum inside the brackets reduces to $|\sum_{k=0}^{11} e^{-i 2\pi m k/12}|^2 = 12^2 \delta_{m0}$, recovering $P_m = \delta_{m0}$. Phase coherence perfectly shields the dark quark, preventing leakage into other mass eigenstates.
On the other hand, for $a_J \to \infty$, all off-diagonal terms vanish ($e^{-a_J d^2} \to 0$ for $d \ge 1$). Only the diagonal term $d = 0$ survives, giving $P_m \to 12 / 12^2 = 1/12$. The system completely decoheres into a statistically uniform mixture over all twelve mass states.

The classical Markov simulation, which replaces off-diagonal coherence with stochastic transition probabilities at every emission step, decoheres more rapidly toward the uniform $1/12$ line. In contrast, QVPS retains history-dependent phase correlations until the final measurement, preserving deeper suppression in the intermediate modes ($m = 3$--$8$) and a visibly sharper coherent interference pattern.

\subsection{Flavor mixing angle}
\label{sec:mass-mixing-angle}
To examine how the degree of flavor mixing affects the shower dynamics, we scan the dimensionless mixing parameter $\xi \in [0, 1]$ in steps of $0.1$, interpolating continuously from the aligned basis ($\xi = 0$, $U = I_{12}$) to the maximal-mixing benchmark ($\xi = 1$, $U = V$).  All other kinematic and gauge parameters are held fixed, and we generate $10^4$ shower events per scan point with both the classical Markovian algorithm and the QVPS quantum simulation.

\begin{figure}[t]
  \centering
  \includegraphics[width=0.48\linewidth]{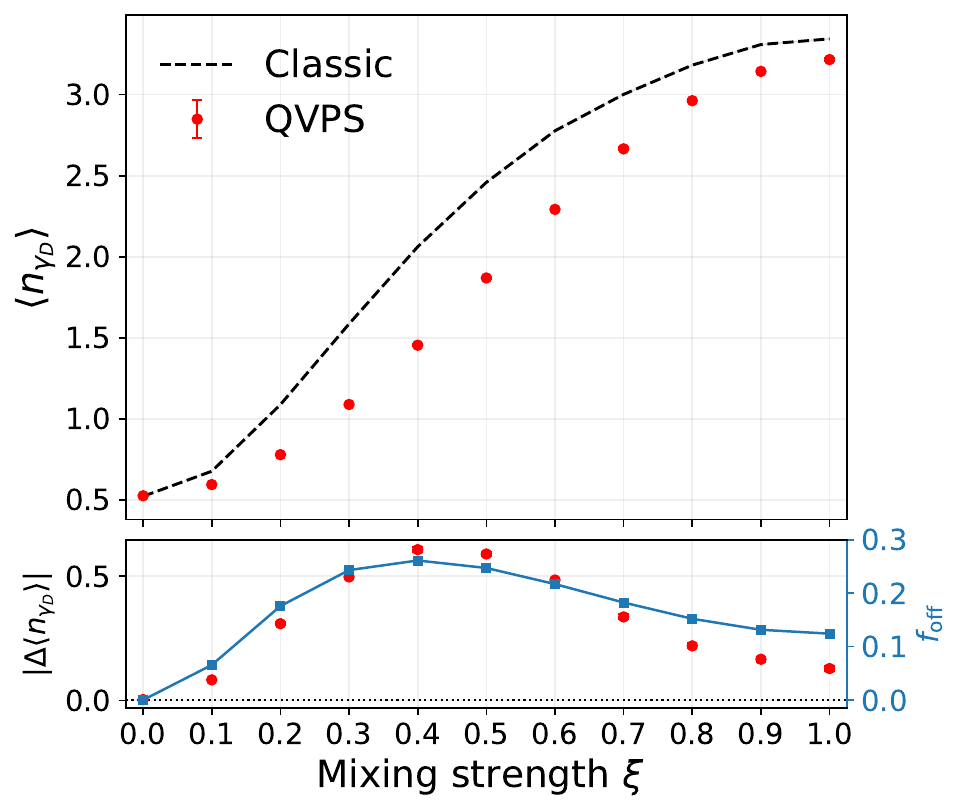}
  \includegraphics[width=0.45\linewidth]{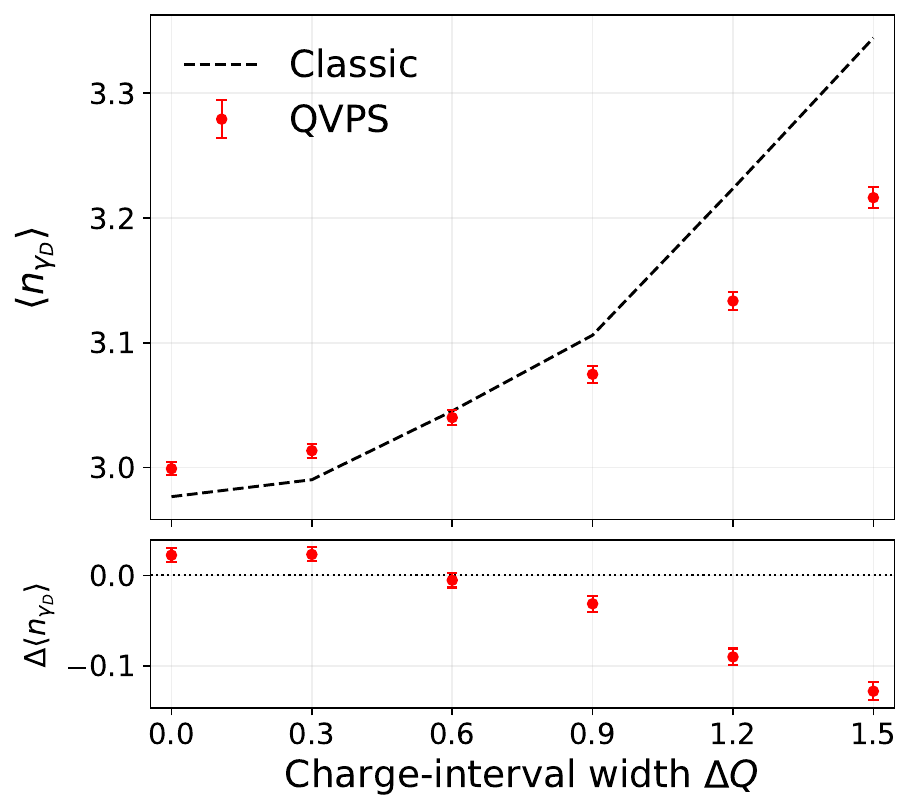}
  \caption{Mean dark-photon multiplicity $\langle n_{\gammad}\rangle$ and quantum--classical differences as functions of the mass-basis mixing strength $\xi$ (left) and the charge-interval width $\Delta Q$ (right). In each panel, the upper plot compares QVPS (red points) with the classical physical-flavor Markov shower (black dashed curve); the lower plot displays the discrepancy $|\Delta\langle n_{\gammad}\rangle|$ (red points). 
In the lower-left plot, the analytic off-diagonal branching fraction $f_{\mathrm{off}}(\xi)$ is superposed. 
  \label{fig:result-mixing-strength-ph}}
\end{figure}

The left panel of Figure~\ref{fig:result-mixing-strength-ph} displays the resulting mean dark-photon multiplicities $\langle n_{\gammad}\rangle$.  In both simulations, $\langle n_{\gammad}\rangle$ increases monotonically with $\xi$, rising from $\approx 0.52$ at $\xi = 0$ to $3.2$--$3.4$ near $\xi = 1$.  This overall enhancement occurs because the initial mass eigenstate of Eq.~\eqref{eq:parameterized-chi0-state} acquires greater support on interaction eigenstates carrying larger $\Uone(1)_D$ charges $Q_k$.
However, the difference between the two simulations, $\Delta\langle n_{\gammad}\rangle = \langle n_{\gammad}\rangle_{\mathrm{QVPS}} - \langle n_{\gammad}\rangle_{\mathrm{Classic}}$, exhibits a distinct, non-monotonic behavior.
Throughout the mixed region ($0 < \xi \le 1$), QVPS systematically produces fewer dark photons than the classical Markov shower, for reasons discussed in the previous subsection. 
The magnitude of the discrepancy $|\Delta\langle n_{\gammad}\rangle|$ reaches a broad maximum near $\xi \approx 0.4$, and $\Delta\langle n_{\gammad}\rangle = 0$ at the aligned endpoint $\xi = 0$. 

The physical origin of this non-monotonic separation lies in how the two frameworks handle transitions between flavor states.  In the mass basis, the dark-photon interaction Hamiltonian is governed by the rotated charge matrix $\widehat{Q}(\xi)$ of Eq.~\eqref{eq:rotated-charge-matrix}, with $\mathbf{Q} = \operatorname{diag}(Q_0, \ldots, Q_{11})$.  Its matrix element $\widehat{Q}_{ji}(\xi) = \langle j_{\mathrm{mass}}| \widehat{Q} |i_{\mathrm{mass}}\rangle$ represents the quantum transition amplitude for the vertex $\chi_i \to \chi_j + \gammad$.  
Upon an accepted emission from state $\chi_0$, if all off-diagonal elements vanish ($\widehat{Q}_{j0} = 0$ for $j \neq 0$), the initial state $\chi_0$ can never change its flavor label, and both classical and quantum showers collapse to a trivial single-flavor evolution with $\Delta\langle n_{\gammad}\rangle = 0$.  Therefore, the divergence between the two descriptions is fundamentally enabled by the availability of flavor-changing paths, measured by the total off-diagonal branching fraction:
\begin{equation}
  f_{\mathrm{off}}(\xi) = \frac{\displaystyle\sum_{j\neq0}|\widehat{Q}_{j0}(\xi)|^2}{\displaystyle\sum_m |\widehat{Q}_{m0}(\xi)|^2}.
  \label{eq:foff-def}
\end{equation}

We can evaluate $f_{\mathrm{off}}(\xi)$ analytically in terms of the charge distribution sampled by the initial mass state.  Expanding the initial state of Eq.~\eqref{eq:parameterized-chi0-state} in the interaction basis, the probability of finding the initial quark in interaction eigenstate $k$ is $p_k(\xi) \equiv |U_{0k}(\xi)|^2$ with $\sum_k p_k = 1$.  Using the unitarity of $U(\xi)$, the denominator and numerator of Eq.~\eqref{eq:foff-def} evaluate to:
\begin{align}
  \sum_m |\widehat{Q}_{m0}|^2 &= \langle 0_{\mathrm{mass}}| \widehat{Q}^\dagger \widehat{Q} |0_{\mathrm{mass}}\rangle = \sum_k p_k(\xi) Q_k^2 \equiv \langle Q^2\rangle_p, \\
  |\widehat{Q}_{00}|^2 &= \left|\langle 0_{\mathrm{mass}}| \widehat{Q} |0_{\mathrm{mass}}\rangle\right|^2 = \left(\sum_k p_k(\xi) Q_k\right)^2 \equiv \langle Q\rangle_p^2.
\end{align}
We superpose the evaluated $f_{\mathrm{off}}$ for each $\xi$ in the lower-left panel of Fig~\ref{fig:result-mixing-strength-ph}.
From which we can observe that the $f_{\mathrm{off}}$ and $|\Delta\langle n_{\gammad}\rangle|$ follow similar tendency, \textit{e.g.} both reach maximum around $\xi \sim 0.4$. 
Note that the final multiplicity difference also depends on the no-emission factors, competing channels, kinematic depletion, and the stepwise shower
construction.

\subsection{Varying the coupling range}

We next fix the mixing matrix at the maximal-mixing endpoint, $\xi=1$, and vary the spread of the twelve interaction-basis charges while keeping their center fixed at $Q_c=1.25$.  The individual charges are parameterized as
\begin{equation}
  Q_k = Q_c + \Delta Q \left[\frac{k}{\Nf-1} - \frac{1}{2}\right],
  \label{eq:charge-spread-def}
\end{equation}
where the charge range $\Delta Q = Q_{\max} - Q_{\min}$ is scanned across $\Delta Q \in \{0,\, 0.3,\, 0.6,\, 0.9,\, 1.2,\, 1.5\}$.  The upper endpoint $\Delta Q = 1.5$ corresponds to our default benchmark interval $[Q_{\min}, Q_{\max}] = [0.5, 2.0]$, whereas $\Delta Q = 0$ collapses the spectrum to twelve degenerate charges $Q_k = 1.25$.  For each scan point, we simulate $10^5$ events with both the classical Markovian shower and the QVPS quantum framework, holding all other kinematic and gauge parameters identical.

Given the charge assignment, we have $\langle Q\rangle = Q_c = 1.25$ and \begin{equation}
  \langle Q^2\rangle = \frac{1}{\Nf}\sum_{k=0}^{\Nf-1} Q_k^2 = Q_c^2 + \frac{\Nf+1}{12(\Nf-1)}\,(\Delta Q)^2.
  \label{eq:Q2-moment}
\end{equation}
Substituting these values into the off-diagonal branching fraction defined in Eq.~\eqref{eq:foff-def}, we obtain
\begin{equation}
  f_{\mathrm{off}}(\Delta Q) = \frac{\langle Q^2\rangle - \langle Q\rangle^2}{\langle Q^2\rangle} = 1 - \frac{Q_c^2}{\langle Q^2\rangle}.
  \label{eq:foff-deltaQ}
\end{equation}
As $\Delta Q$ increases from $0$ to $1.5$, the $f_{\mathrm{off}}$  grows monotonically from $0$ to $\approx 0.12$.  In the degenerate limit $\Delta Q=0$, we have $\mathbf{Q}=Q_c I_{12}$ and $U\mathbf{Q}U^\dagger=Q_c I_{12}$, making the dark-photon interaction diagonal in every basis. Moreover, the flavor transitions are forbidden and both showers collapse to identical single-flavor evolution.

The mean dark photon multiplicity with varying $\Delta Q$ are presented in the right panel of Fig.~\ref{fig:result-mixing-strength-ph}. 
Both the classical and quantum algorithms exhibit a monotonic increase in the mean dark-photon multiplicity $\langle n_{\gammad}\rangle$ as $\Delta Q$ widens. This common upward trend is primarily driven by the growth of the effective coupling scale. Because the average squared charge $\langle Q^2\rangle$ increases with $\Delta Q$ according to Eq.~\eqref{eq:Q2-moment}, the integrated emission rate across the ensemble is naturally enhanced in both frameworks. 
On the other hand, the classical shower exhibits a steeper multiplicity climb than the quantum shower, with a broader charge spectrum. As illustrated in the lower panel, the two showers diverge systematically into the negative territory, the deficit reaching about $-0.13$ at $\Delta Q=1.5$, of the same size as the reduction found at maximal mixing in Fig.~\ref{fig:result-emission-counts}.
Such a widening discrepancy is attributed to the coherent feedback effect discussed in Sec.~\ref{sec:classical-vs-quantum}:  for small $\Delta Q$ the angles $\theta_{\gammad,k}^{(j)}$ are nearly identical across all $k$ and the veto acts almost uniformly, whereas for larger $\Delta Q$ the charge contrast grows and the suppression of the larger-charge components in each survival interval, $\cos(\theta_{\gammad,k}^{(j)}/2)\approx 1-p_{j,k}/2$, becomes increasingly effective.  
This continuous, coherent filtering dynamically shifts the emitter's statevector toward lower-$Q_k$ components along no-emission histories, creating an effective negative feedback that suppresses subsequent emissions in the quantum shower; its magnitude can be tracked by the available flavor-changing phase space $f_{\mathrm{off}}(\Delta Q)$ of Eq.~\eqref{eq:foff-deltaQ}.

Crucially, the degenerate endpoint $\Delta Q = 0$ shows a small positive residual offset, {\textit i.e.} $\Delta\langle n_{\gammad}\rangle \sim 0.02 $. 
Because all twelve charges are identical at $\Delta Q = 0$, the veto is strictly flavor-blind and no quantum interference can occur.  This small offset instead stems from a subtle finite-step algorithmic difference in how the two frameworks map integrated splitting rates to discrete emission probabilities.

At each evolution step $j$, the QVPS framework implements a linearized acceptance probability directly proportional to the integrated splitting density, giving for the first step
\begin{equation}
  p_{\gammad,\mathrm{QVPS}}^{(1)} = \int dz\, g_{\chi\to\chi\gammad}^{(1)}(t_1^2, z).
\end{equation}
In contrast, the classical shower first evaluates whether any emission occurs using the finite-step exponential Sudakov factor $1 - \Delta(t_1^2; t_0^2) = 1 - \exp\left(-p_{\mathrm{tot},\mathrm{QVPS}}^{(1)}\right)$, where $p_{\mathrm{tot},\mathrm{QVPS}}^{(1)} \equiv \sum_a p_{a,\mathrm{QVPS}}^{(1)}$ is the total first-step emission probability summed over all competing channels ($\chi\to\chi\gammad$ and $\chi\to\chi g$).  Conditioned on an emission occurring, the dark-photon channel is selected according to its relative rate, leading to
\begin{equation}
  p_{\gammad,\mathrm{Classic}}^{(1)} = \left(1 - e^{-p_{\mathrm{tot},\mathrm{QVPS}}^{(1)}}\right) \frac{p_{\gammad,\mathrm{QVPS}}^{(1)}}{p_{\mathrm{tot},\mathrm{QVPS}}^{(1)}} = p_{\gammad,\mathrm{QVPS}}^{(1)} \left[1 - \frac{1}{2}p_{\mathrm{tot},\mathrm{QVPS}}^{(1)} + \mathcal{O}\!\left(\left(p_{\mathrm{tot},\mathrm{QVPS}}^{(1)}\right)^2\right)\right].
  \label{eq:classic-first-step-photon-probability}
\end{equation}
Consequently, at finite step size $(-\ln r)$, $p_{\gammad,\mathrm{QVPS}}^{(1)}$ exceeds $p_{\gammad,\mathrm{Classic}}^{(1)}$ by $\frac{1}{2} p_{\gammad,\mathrm{QVPS}}^{(1)} p_{\mathrm{tot},\mathrm{QVPS}}^{(1)}$ at leading order.  

\section{Conclusion}
\label{sec:conclusion}
In this work, we have developed and validated a QVPS framework for a finite-mass dark sector within the conformal window, governed by the gauge group $\mathrm{SU}(3)_D \times \mathrm{U}(1)_D$ with $N_C = 3$ and $N_F = 12$. The non-Abelian sector incorporates the exact two-loop running coupling via the Lambert-$W$ formulation, naturally capturing the trajectory toward an infrared fixed point ($\alpha_* \approx 0.754$) and the subsequent departure from conformality governed by the dark-quark mass threshold $M_q$. The Abelian sector introduces twelve interaction eigenstates with non-degenerate charges, while a continuous unitary rotation connects the interaction basis to the mass basis where initial states are prepared and final observables are measured.

Our numerical simulations and analytic derivations establish several key physical findings:
\begin{itemize}
    \item \textbf{EEC Angular Scaling:} The all-pairs energy-energy correlation shows approximate linearity in double-logarithmic coordinates over an intermediate angular range ($10^{-3} \lesssim R_L \lesssim 10^{-1}$), suggesting power-law scaling. This behavior is qualitatively compatible with scaling in a near-conformal regime. The power-law scaling region is restricted at small angles by the mass cutoff $t^2_{\mathrm{cut}} = 4M_q^2$.
    
    \item \textbf{Enhanced Zero-emission Rate:} The charge-dependent quantum veto gate dynamically imprints the splitting probability onto the flavor statevector without collapsing it. Along no-emission histories, the statevector is coherently filtered toward lower-charge components. This quantum negative feedback yields a marked enhancement in the zero-emission fraction ($14.54\%$ in QVPS compared to $3.38\%$ classically) and reduces the mean dark-photon multiplicity $\langle n_{\gamma_D} \rangle$, while leaving the flavor-blind dark-gluon radiation largely unaffected.
    
    \item \textbf{U-Shaped Mass Spectrum:} Coherent flavor evolution preserves relative phases across successive splittings, in contrast to the stochastic flavor projection of classical showers. When projected back into the mass basis at the end of the shower, this phase coherence creates an interference pattern characterized by a U-shaped mass spectrum: enhancements at $m=0$ and $m=11$ and destructive interference across intermediate states ($m \in [3, 8]$). 
    
    \item \textbf{Parametric Sensitivity:} Continuous scans over the mixing parameter $\xi$ and charge spread $\Delta Q$ demonstrate that the quantum--classical discrepancy $|\Delta \langle n_{\gamma_D} \rangle|$ directly tracks the available off-diagonal flavor-changing phase space $f_{\mathrm{off}}$, vanishing in the aligned limit ($\xi = 0$ or $\Delta Q = 0$) and peaking near $\xi \approx 0.4$.
\end{itemize}

While vector-to-fermion pair productions ($g \to \chi\bar{\chi}$, $\gamma_D \to \chi\bar{\chi}$) were omitted in the full numerical shower due to the exponential statevector memory overhead on classical simulators, the underlying multi-register entanglement circuits have been formulated. 
Future work will focus on incorporating full kinematic recoil schemes, extending the framework toward hadronization models, and testing the algorithm on physical quantum hardware to explore dark-jet substructure at future colliders.

\begin{acknowledgments}
This work was supported by the National Natural Science Foundation of China under grant Nos. 12335002, 12275185, 12575106, 12675105, 12675138, the Natural Science Foundation of Sichuan Province under grant No. 2026NSFSC0034 and Basic Scientific Research Project of the Education Department of Liaoning Province under Grant No.LJ212610165009. We acknowledge the use of GPT-Sol and DeepSeek for assistance with code development and manuscript preparation. The authors take full responsibility for the content.

\end{acknowledgments}

\begin{appendices}
\section*{Appendices}
\section{Validation of the simulation framework}
\label{app:table2-benchmark}
To validate the implementation of the parton shower within our quantum veto framework, we benchmark our simulation against the radiating dark matter model of Ref.~\cite{Buschmann:2015awa}. 

We simplify the full framework by turning off the $\SU(3)_D$ sector, flavor mixing, and pair production ($A'\to\chi\bar\chi$), retaining a single fermion species emitting a massive dark photon via $\chi \to \chi A'$ with coupling $\alpha_{A'} = 0.2$, exactly as in the analytic and Pythia comparisons in the reference. The multi-qubit flavor register is thus omitted, and the quantum veto operation reduces to a single $R_Y$ rotation on the emission qubit at each evolution step. 

Following Ref.~\cite{Buschmann:2015awa}, the reference process is $\chi\bar{\chi}$ production at $\sqrt{\hat{s}} = 1~\mathrm{TeV}$, giving an initial radiator energy $E_0 = 500~\mathrm{GeV}$. The reduced quantum-veto shower uses the angular-ordered evolution variable of Sec.~\ref{sec:framework}, $t^2\equiv\widetilde q^{\,2}=E_{\mathrm{parent}}^2(\Delta R_{\mathrm{split}})^2$ in the small-angle approximation. 
In contrast with Eq.~(\ref{eq:z-limits}), the mass-dependent sampling interval and infrared endpoint are
\begin{equation}
  z_-(t^2)=\frac{m_\chi}{\sqrt{t^2}},\qquad
  z_+(t^2)=1-\frac{m_{A'}}{\sqrt{t^2}},\qquad
  t_{\mathrm{cut}}^2=(m_\chi+m_{A'})^2,
\end{equation}
since $m_\chi$ and $m_{A^\prime}$ differ dramatically for the benchmark points in the reference. 
At each step, $z$ is proposed from an overestimate proportional to $1/(1-z)$, while the fermion splitting kernel is $P_{\chi\to\chi}(z)=(1+z^2)/(1-z)$. The emission-qubit $R_Y$ rotation, measurement, and reset implement the quantum-veto decision. After an accepted $\chi\to\chi A'$ splitting at $t_{\mathrm{split}}^2$, the continuing fermion starts its subsequent evolution at $z^2t_{\mathrm{split}}^2$; the emitted dark photon does not radiate in this reduced benchmark. The observables used for validation include the average dark-photon multiplicity for the pair $2\langle n_{A'}\rangle$, the final fermion energy fraction $\langle X\rangle = \langle E_{\mathrm{final}}/E_0\rangle$, and the average photon energy fraction $\langle Z\rangle$, which satisfy the recoil conservation relation $\langle X\rangle + \langle n_{A'}\rangle \langle Z\rangle = 1$.

\begin{figure}[htbp]
  \centering
  \includegraphics[width=1\linewidth]{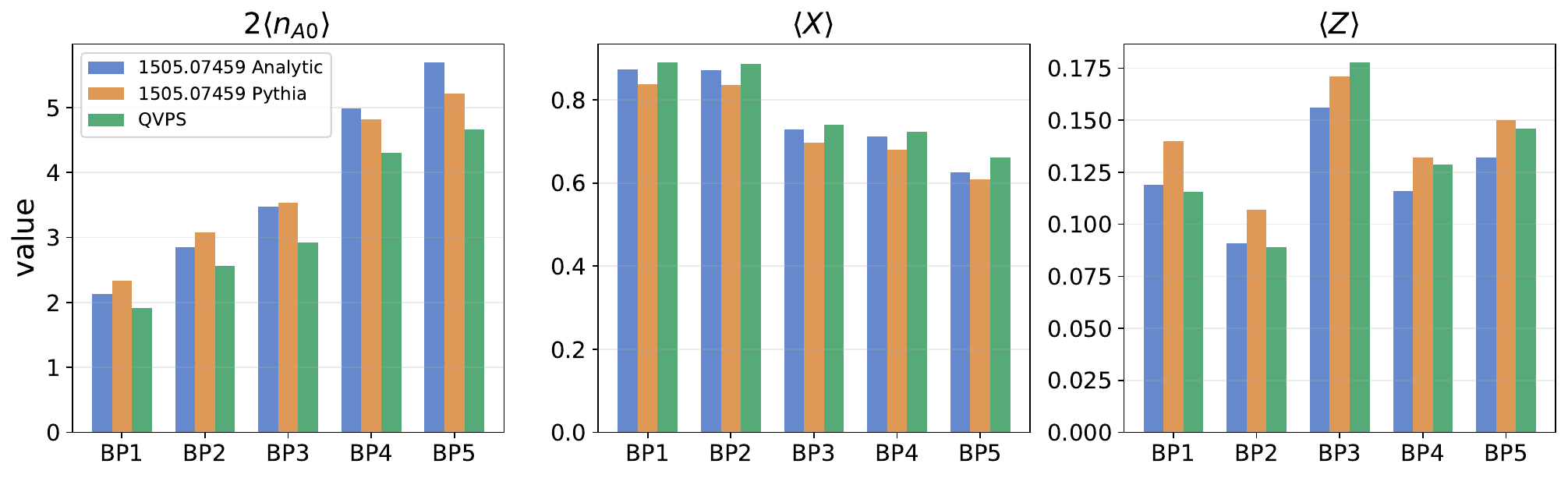}
  \caption{Comparison of the reduced quantum-veto benchmark with Table~II of Ref.~\cite{Buschmann:2015awa}. BP1--BP5 follow the mass points defined in the text. Blue and orange bars show the semi-analytic and Pythia results from the reference, while green bars show the reduced QVPS results.
  \label{fig:benchmark-comparison}}
\end{figure}

We simulate the five benchmark points $(m_\chi, m_{A'})$ in $\mathrm{GeV}$ from Table~II of the reference:
\[
  \mathrm{BP1}\ (50,1.5),\quad
  \mathrm{BP2}\ (50,0.4),\quad
  \mathrm{BP3}\ (4,1.5),\quad
  \mathrm{BP4}\ (4,0.4),\quad
  \mathrm{BP5}\ (0.4,0.4).
\]
For each point, 1000 events are generated. Figure~\ref{fig:benchmark-comparison} compares our results with the semi-analytic and Pythia~8 values quoted in Ref.~\cite{Buschmann:2015awa}.
The QVPS benchmark shows good overall agreement with both reference calculations across all mass hierarchies. The modest reduction in $2\langle n_{A'}\rangle$ across the five benchmark points is attributed to the difference in the shower evolution variables.

\end{appendices}

\bibliographystyle{jhep}
\bibliography{jhep}

@article{Gardi:1998qr,
    author = "Gardi, Einan and Grunberg, Georges and Karliner, Marek",
    title = "{Can the QCD running coupling have a causal analyticity structure?}",
    eprint = "hep-ph/9806462",
    archivePrefix = "arXiv",
    reportNumber = "TAUP-2503-98",
    doi = "10.1088/1126-6708/1998/07/007",
    journal = "JHEP",
    volume = "07",
    pages = "007",
    year = "1998"
}

@article{Gardi:1998rf,
    author = "Gardi, Einan and Karliner, Marek",
    title = "{Relations between observables and the infrared fixed point in QCD}",
    eprint = "hep-ph/9802218",
    archivePrefix = "arXiv",
    reportNumber = "TAUP-2472-98",
    doi = "10.1016/S0550-3213(98)00392-7",
    journal = "Nucl. Phys. B",
    volume = "529",
    pages = "383--423",
    year = "1998"
}

@article{Magradze:1999um,
    author = "Magradze, B. A.",
    title = "{Analytic approach to perturbative QCD}",
    eprint = "hep-ph/9911456",
    archivePrefix = "arXiv",
    reportNumber = "RMI-99-06",
    doi = "10.1142/S0217751X00001115",
    journal = "Int. J. Mod. Phys. A",
    volume = "15",
    pages = "2715--2734",
    year = "2000"
}

@article{Caswell:1974gg,
    author = "Caswell, William E.",
    title = "{Asymptotic Behavior of Nonabelian Gauge Theories to Two Loop Order}",
    reportNumber = "PRINT-74-1058 (PRINCETON)",
    doi = "10.1103/PhysRevLett.33.244",
    journal = "Phys. Rev. Lett.",
    volume = "33",
    pages = "244",
    year = "1974"
}

@article{Banks:1981nn,
    author = "Banks, Tom and Zaks, A.",
    title = "{On the Phase Structure of Vector-Like Gauge Theories with Massless Fermions}",
    reportNumber = "TAUP-944-81",
    doi = "10.1016/0550-3213(82)90035-9",
    journal = "Nucl. Phys. B",
    volume = "196",
    pages = "189--204",
    year = "1982"
}

@article{Prosperi:2006hx,
    author = "Prosperi, G. M. and Raciti, M. and Simolo, C.",
    title = "{On the running coupling constant in QCD}",
    eprint = "hep-ph/0607209",
    archivePrefix = "arXiv",
    doi = "10.1016/j.ppnp.2006.09.001",
    journal = "Prog. Part. Nucl. Phys.",
    volume = "58",
    pages = "387--438",
    year = "2007"
}

@article{Kulkarni:2025rsl,
    author = "Kulkarni, Suchita and Lockyer, Joshua and Strassler, Matthew J.",
    title = "{On the simulation of hidden parton showers in the conformal window}",
    eprint = "2502.18566",
    archivePrefix = "arXiv",
    primaryClass = "hep-ph",
    doi = "10.1007/JHEP09(2025)150",
    journal = "JHEP",
    volume = "09",
    pages = "150",
    year = "2025"
}

@article{Bauer:2023ujy,
    author = "Bauer, Christian W. and Chigusa, So and Yamazaki, Masahito",
    title = "{Quantum parton shower with kinematics}",
    eprint = "2310.19881",
    archivePrefix = "arXiv",
    primaryClass = "hep-ph",
    doi = "10.1103/PhysRevA.109.032432",
    journal = "Phys. Rev. A",
    volume = "109",
    number = "3",
    pages = "032432",
    year = "2024"
}

@article{Collins:1989bt,
    author = "Collins, John C.",
    title = "{Sudakov form-factors}",
    eprint = "hep-ph/0312336",
    archivePrefix = "arXiv",
    doi = "10.1142/9789814503266_0006",
    journal = "Adv. Ser. Direct. High Energy Phys.",
    volume = "5",
    pages = "573--614",
    year = "1989"
}

@inproceedings{Hoche:2014rga,
    author = {H{\"o}che, Stefan},
    title = "{Introduction to parton-shower event generators}",
    booktitle = "{Theoretical Advanced Study Institute in Elementary Particle Physics}: {Journeys Through the Precision Frontier: Amplitudes for Colliders}",
    eprint = "1411.4085",
    archivePrefix = "arXiv",
    primaryClass = "hep-ph",
    reportNumber = "SLAC-PUB-16160",
    doi = "10.1142/9789814678766_0005",
    pages = "235--295",
    year = "2015"
}

@article{Nagy:2017ggp,
    author = "Nagy, Zoltan and Soper, Davison E.",
    title = "{What is a parton shower?}",
    eprint = "1705.08093",
    archivePrefix = "arXiv",
    primaryClass = "hep-ph",
    reportNumber = "DESY-17-063, DESY 17-063",
    doi = "10.1103/PhysRevD.98.014034",
    journal = "Phys. Rev. D",
    volume = "98",
    number = "1",
    pages = "014034",
    year = "2018"
}

@article{Papaefstathiou:2024qlg,
    author = "Papaefstathiou, Andreas",
    title = "{$\texttt{Pyresias}$: How To Write a Toy Parton Shower}",
    eprint = "2406.03528",
    archivePrefix = "arXiv",
    primaryClass = "hep-ph",
    reportNumber = "MCNET-24-08",
    month = "6",
    year = "2024"
}

@article{Borrello:2026bfr,
    author = "Borrello, Matteo and Costa, Marco and Redigolo, Diego and Tammaro, Michele",
    title = "{Probing Neutrino Compositeness with Invisible and Displaced Signals}",
    eprint = "2604.14282",
    archivePrefix = "arXiv",
    primaryClass = "hep-ph",
    month = "4",
    year = "2026"
}

@article{Kim:2026mfk,
    author = "Kim, Taehee and Lee, Joon-Bin and Masouminia, M. R. and Seymour, Michael H. and Yang, Un-ki",
    title = "{Searching for New Physics Inside Jets with the Herwig 7 Generalised Parton Shower}",
    eprint = "2604.00617",
    archivePrefix = "arXiv",
    primaryClass = "hep-ph",
    reportNumber = "CERN-TH-2026-074, MCnet-26-05, IPPP-26-15",
    month = "4",
    year = "2026"
}

@article{Chigusa:2022act,
    author = "Chigusa, So and Yamazaki, Masahito",
    title = "{Quantum simulations of dark sector showers}",
    eprint = "2204.12500",
    archivePrefix = "arXiv",
    primaryClass = "hep-ph",
    reportNumber = "KEK-TH-2417",
    doi = "10.1016/j.physletb.2022.137466",
    journal = "Phys. Lett. B",
    volume = "834",
    pages = "137466",
    year = "2022"
}

@article{Strassler:2006im,
    author = "Strassler, Matthew J. and Zurek, Kathryn M.",
    title = "{Echoes of a hidden valley at hadron colliders}",
    eprint = "hep-ph/0604261",
    archivePrefix = "arXiv",
    doi = "10.1016/j.physletb.2007.06.055",
    journal = "Phys. Lett. B",
    volume = "651",
    pages = "374--379",
    year = "2007"
}

@article{Han:2007ae,
    author = "Han, Tao and Si, Zongguo and Zurek, Kathryn M. and Strassler, Matthew J.",
    title = "{Phenomenology of hidden valleys at hadron colliders}",
    eprint = "0712.2041",
    archivePrefix = "arXiv",
    primaryClass = "hep-ph",
    reportNumber = "MADPH-07-1502, SDU-HEP-071201, RUNHETC-2007-31",
    doi = "10.1088/1126-6708/2008/07/008",
    journal = "JHEP",
    volume = "07",
    pages = "008",
    year = "2008"
}

@article{Gori:2022vri,
    author = "Gori, Stefania and others",
    title = "{Dark Sector Physics at High-Intensity Experiments}",
    eprint = "2209.04671",
    archivePrefix = "arXiv",
    primaryClass = "hep-ph",
    reportNumber = "FERMILAB-PUB-22-672-SCD-T",
    month = "9",
    year = "2022"
}

@article{Cohen:2017pzm,
    author = "Cohen, Timothy and Lisanti, Mariangela and Lou, Hou Keong and Mishra-Sharma, Siddharth",
    title = "{LHC Searches for Dark Sector Showers}",
    eprint = "1707.05326",
    archivePrefix = "arXiv",
    primaryClass = "hep-ph",
    reportNumber = "PUPT-2529",
    doi = "10.1007/JHEP11(2017)196",
    journal = "JHEP",
    volume = "11",
    pages = "196",
    year = "2017"
}

@article{Cohen:2020afv,
    author = "Cohen, Timothy and Doss, Joel and Freytsis, Marat",
    title = "{Jet Substructure from Dark Sector Showers}",
    eprint = "2004.00631",
    archivePrefix = "arXiv",
    primaryClass = "hep-ph",
    doi = "10.1007/JHEP09(2020)118",
    journal = "JHEP",
    volume = "09",
    pages = "118",
    year = "2020"
}

@article{Knapen:2021eip,
    author = "Knapen, Simon and Shelton, Jessie and Xu, Dong",
    title = "{Perturbative benchmark models for a dark shower search program}",
    eprint = "2103.01238",
    archivePrefix = "arXiv",
    primaryClass = "hep-ph",
    doi = "10.1103/PhysRevD.103.115013",
    journal = "Phys. Rev. D",
    volume = "103",
    number = "11",
    pages = "115013",
    year = "2021"
}

@article{Albouy:2022cin,
    author = "Albouy, Guillaume and others",
    title = "{Theory, phenomenology, and experimental avenues for dark showers: a Snowmass 2021 report}",
    eprint = "2203.09503",
    archivePrefix = "arXiv",
    primaryClass = "hep-ph",
    reportNumber = "FERMILAB-PUB-22-303-PPD-SCD-T",
    doi = "10.1140/epjc/s10052-022-11048-8",
    journal = "Eur. Phys. J. C",
    volume = "82",
    number = "12",
    pages = "1132",
    year = "2022"
}

@article{Schwaller:2015gea,
    author = "Schwaller, Pedro and Stolarski, Daniel and Weiler, Andreas",
    title = "{Emerging Jets}",
    eprint = "1502.05409",
    archivePrefix = "arXiv",
    primaryClass = "hep-ph",
    reportNumber = "CERN-PH-TH-2015-031, DESY-15-026",
    doi = "10.1007/JHEP05(2015)059",
    journal = "JHEP",
    volume = "05",
    pages = "059",
    year = "2015"
}

@article{CMS:2018bvr,
    author = "Sirunyan, Albert M and others",
    collaboration = "CMS",
    title = "{Search for new particles decaying to a jet and an emerging jet}",
    eprint = "1810.10069",
    archivePrefix = "arXiv",
    primaryClass = "hep-ex",
    reportNumber = "CMS-EXO-18-001, CERN-EP-2018-255",
    doi = "10.1007/JHEP02(2019)179",
    journal = "JHEP",
    volume = "02",
    pages = "179",
    year = "2019"
}

@article{Carrasco:2023loy,
    author = "Carrasco, Juliana and Zurita, Jos{\'e}",
    title = "{Emerging jet probes of strongly interacting dark sectors}",
    eprint = "2307.04847",
    archivePrefix = "arXiv",
    primaryClass = "hep-ph",
    doi = "10.1007/JHEP01(2024)034",
    journal = "JHEP",
    volume = "01",
    pages = "034",
    year = "2024"
}

@article{ATLAS:2025lfx,
    author = "Aad, Georges and others",
    collaboration = "ATLAS",
    title = "{Search for emerging jets in $pp$ collisions at $\sqrt{s} = 13$ TeV with the ATLAS experiment}",
    eprint = "2510.12347",
    archivePrefix = "arXiv",
    primaryClass = "hep-ex",
    reportNumber = "CERN-EP-2025-225",
    month = "10",
    year = "2025"
}

@article{Carrasco:2025bct,
    author = "Carrasco, Juliana and Kulkarni, Suchita and Liu, Wei and Lockyer, Joshua and Zurita, Jose",
    title = "{Semi-visible emerging jets}",
    eprint = "2511.02918",
    archivePrefix = "arXiv",
    primaryClass = "hep-ph",
    month = "11",
    year = "2025"
}

@article{Cohen:2015toa,
    author = "Cohen, Timothy and Lisanti, Mariangela and Lou, Hou Keong",
    title = "{Semivisible Jets: Dark Matter Undercover at the LHC}",
    eprint = "1503.00009",
    archivePrefix = "arXiv",
    primaryClass = "hep-ph",
    doi = "10.1103/PhysRevLett.115.171804",
    journal = "Phys. Rev. Lett.",
    volume = "115",
    number = "17",
    pages = "171804",
    year = "2015"
}

@article{Beauchesne:2017yhh,
    author = "Beauchesne, Hugues and Bertuzzo, Enrico and Grilli Di Cortona, Giovanni and Tabrizi, Zahra",
    title = "{Collider phenomenology of Hidden Valley mediators of spin 0 or 1/2 with semivisible jets}",
    eprint = "1712.07160",
    archivePrefix = "arXiv",
    primaryClass = "hep-ph",
    doi = "10.1007/JHEP08(2018)030",
    journal = "JHEP",
    volume = "08",
    pages = "030",
    year = "2018"
}

@article{CMS:2021dzg,
    author = "Tumasyan, Armen and others",
    collaboration = "CMS",
    title = "{Search for resonant production of strongly coupled dark matter in proton-proton collisions at 13 TeV}",
    eprint = "2112.11125",
    archivePrefix = "arXiv",
    primaryClass = "hep-ex",
    reportNumber = "CMS-EXO-19-020, CERN-EP-2021-252",
    doi = "10.1007/JHEP06(2022)156",
    journal = "JHEP",
    volume = "06",
    pages = "156",
    year = "2022"
}

@article{Liu:2024rbe,
    author = "Liu, Bingxuan and Pedro, Kevin",
    title = "{Semi-visible jets + X: illuminating dark showers with radiation}",
    eprint = "2409.04741",
    archivePrefix = "arXiv",
    primaryClass = "hep-ph",
    reportNumber = "FERMILAB-PUB-24-0563-CSAID-PPD",
    doi = "10.1007/JHEP12(2024)105",
    journal = "JHEP",
    volume = "12",
    pages = "105",
    year = "2024"
}

@article{Buckley:2025hty,
    author = "Buckley, Andy and Butterworth, Jon and Corpe, Louie and Doglioni, Caterina and Kar, Deepak and Prat, Clarisse and Sinha, Sukanya and Wilson-Edwards, Danielle",
    title = "{Probing the sensitivity of semi-visible jets to current LHC measurements using the CONTUR toolkit}",
    eprint = "2502.11237",
    archivePrefix = "arXiv",
    primaryClass = "hep-ph",
    month = "2",
    year = "2025"
}

@article{ATLAS:2025kuz,
    author = "Aad, Georges and others",
    collaboration = "ATLAS",
    title = "{Search for new physics in final states with semivisible jets or anomalous signatures using the ATLAS detector}",
    eprint = "2505.01634",
    archivePrefix = "arXiv",
    primaryClass = "hep-ex",
    reportNumber = "CERN-EP-2025-101",
    doi = "10.1103/44zp-mh1q",
    journal = "Phys. Rev. D",
    volume = "112",
    number = "1",
    pages = "012021",
    year = "2025"
}

@article{Cheng:2021kjg,
    author = "Cheng, Hsin-Chia and Li, Lingfeng and Salvioni, Ennio",
    title = "{A theory of dark pions}",
    eprint = "2110.10691",
    archivePrefix = "arXiv",
    primaryClass = "hep-ph",
    reportNumber = "CERN-TH-2021-150",
    doi = "10.1007/JHEP01(2022)122",
    journal = "JHEP",
    volume = "01",
    pages = "122",
    year = "2022"
}

@article{Born:2023vll,
    author = "Born, Susan and Karur, Rohith and Knapen, Simon and Shelton, Jessie",
    title = "{Scouting for dark showers at CMS and LHCb}",
    eprint = "2303.04167",
    archivePrefix = "arXiv",
    primaryClass = "hep-ph",
    doi = "10.1103/PhysRevD.108.035034",
    journal = "Phys. Rev. D",
    volume = "108",
    number = "3",
    pages = "035034",
    year = "2023"
}

@article{Cheng:2024hvq,
    author = "Cheng, Hsin-Chia and Jiang, Xu-Hui and Li, Lingfeng and Salvioni, Ennio",
    title = "{Dark showers from Z-dark Z' mixing}",
    eprint = "2401.08785",
    archivePrefix = "arXiv",
    primaryClass = "hep-ph",
    doi = "10.1007/JHEP04(2024)081",
    journal = "JHEP",
    volume = "04",
    pages = "081",
    year = "2024"
}

@article{Cheng:2024aco,
    author = "Cheng, Hsin-Chia and Jiang, Xu-Hui and Li, Lingfeng",
    title = "{Phenomenology of electroweak portal dark showers: high energy direct probes and low energy complementarity}",
    eprint = "2408.13304",
    archivePrefix = "arXiv",
    primaryClass = "hep-ph",
    doi = "10.1007/JHEP01(2025)149",
    journal = "JHEP",
    volume = "01",
    pages = "149",
    year = "2025"
}

@article{Liebersbach:2024kzc,
    author = "Liebersbach, Samuel and Sandick, Pearl and Shiferaw, Abel and Zhao, Yue",
    title = "{Exploring the hidden valley at MATHUSLA}",
    eprint = "2408.07756",
    archivePrefix = "arXiv",
    primaryClass = "hep-ph",
    doi = "10.1016/j.nuclphysb.2025.117050",
    journal = "Nucl. Phys. B",
    volume = "1018",
    pages = "117050",
    year = "2025"
}

@article{Liu:2025bbc,
    author = "Liu, Wei and Lockyer, Joshua and Kulkarni, Suchita",
    title = "{Hidden valley scenario sensitivity in the CMS muon end cap detector}",
    eprint = "2505.03058",
    archivePrefix = "arXiv",
    primaryClass = "hep-ph",
    doi = "10.1103/j4dk-qfy4",
    journal = "Phys. Rev. D",
    volume = "112",
    number = "7",
    pages = "075029",
    year = "2025"
}

@article{CMS:2025mym,
    collaboration = "CMS",
    title = "{Search for low-mass hidden valley dark showers with displaced dimuons in proton-proton collisions at $\sqrt{s}=13~\mathrm{TeV}$}",
    journal = "CMS-PAS-EXO-24-008",
    year = "2025"
}

@article{Chen:2025btv,
    author = "Chen, Wanyun and Lu, Chih-Ting and Shen, Hanxin",
    title = "{Probing gluons-enriched dark jets from Higgs boson exotic decays at the LHC}",
    eprint = "2512.24814",
    archivePrefix = "arXiv",
    primaryClass = "hep-ph",
    month = "12",
    year = "2025"
}

@article{Harnik:2008ax,
    author = "Harnik, Roni and Wizansky, Tommer",
    title = "{Signals of New Physics in the Underlying Event}",
    eprint = "0810.3948",
    archivePrefix = "arXiv",
    primaryClass = "hep-ph",
    reportNumber = "SLAC-PUB-13483, SU-ITP-26-08",
    doi = "10.1103/PhysRevD.80.075015",
    journal = "Phys. Rev. D",
    volume = "80",
    pages = "075015",
    year = "2009"
}

@article{Knapen:2016hky,
    author = "Knapen, Simon and Pagan Griso, Simone and Papucci, Michele and Robinson, Dean J.",
    title = "{Triggering Soft Bombs at the LHC}",
    eprint = "1612.00850",
    archivePrefix = "arXiv",
    primaryClass = "hep-ph",
    doi = "10.1007/JHEP08(2017)076",
    journal = "JHEP",
    volume = "08",
    pages = "076",
    year = "2017"
}

@article{CMS:2024nca,
    author = "Hayrapetyan, Aram and others",
    collaboration = "CMS",
    title = "{Search for Soft Unclustered Energy Patterns in Proton-Proton Collisions at 13~TeV}",
    eprint = "2403.05311",
    archivePrefix = "arXiv",
    primaryClass = "hep-ex",
    reportNumber = "CMS-EXO-23-002, CERN-EP-2024-054",
    doi = "10.1103/PhysRevLett.133.191902",
    journal = "Phys. Rev. Lett.",
    volume = "133",
    number = "19",
    pages = "191902",
    year = "2024"
}

@article{Asadi:2025vfr,
    author = "Asadi, Pouya and Batz, Austin and Bernreuther, Elias and Costa, Marco and Homiller, Samuel and Kribs, Graham D.",
    title = "{Stopping dark mesons in their tracks with long-lived particle and resonant signatures}",
    eprint = "2507.13430",
    archivePrefix = "arXiv",
    primaryClass = "hep-ph",
    doi = "10.1007/JHEP03(2026)234",
    journal = "JHEP",
    volume = "03",
    pages = "234",
    year = "2026"
}

@article{Carloni:2010tw,
    author = "Carloni, Lisa and Sjostrand, Torbjorn",
    title = "{Visible Effects of Invisible Hidden Valley Radiation}",
    eprint = "1006.2911",
    archivePrefix = "arXiv",
    primaryClass = "hep-ph",
    reportNumber = "LU-TP-10-17, MCNET-10-11",
    doi = "10.1007/JHEP09(2010)105",
    journal = "JHEP",
    volume = "09",
    pages = "105",
    year = "2010"
}

@article{Carloni:2011kk,
    author = "Carloni, Lisa and Rathsman, Johan and Sjostrand, Torbjorn",
    title = "{Discerning Secluded Sector gauge structures}",
    eprint = "1102.3795",
    archivePrefix = "arXiv",
    primaryClass = "hep-ph",
    reportNumber = "LU-TP-11-09, MCNET-11-06",
    doi = "10.1007/JHEP04(2011)091",
    journal = "JHEP",
    volume = "04",
    pages = "091",
    year = "2011"
}

@article{Bierlich:2022pfr,
    author = "Bierlich, Christian and others",
    title = "{A comprehensive guide to the physics and usage of PYTHIA 8.3}",
    eprint = "2203.11601",
    archivePrefix = "arXiv",
    primaryClass = "hep-ph",
    reportNumber = "LU-TP 22-16, MCNET-22-04, FERMILAB-PUB-22-227-SCD",
    doi = "10.21468/SciPostPhysCodeb.8",
    journal = "SciPost Phys. Codeb.",
    volume = "2022",
    pages = "8",
    year = "2022"
}

@article{Kulkarni:2024okx,
    author = {Kulkarni, Suchita and Masouminia, M. R. and Pl{\"a}tzer, Simon and Stafford, Dominic},
    title = "{Dark sector showers and hadronisation in Herwig 7}",
    eprint = "2408.10044",
    archivePrefix = "arXiv",
    primaryClass = "hep-ph",
    reportNumber = "IPPP/24/54, IPPP/24/54; PUBDB-2024-06237",
    doi = "10.1140/epjc/s10052-024-13587-8",
    journal = "Eur. Phys. J. C",
    volume = "84",
    number = "11",
    pages = "1210",
    year = "2024"
}

@article{Curtin:2022tou,
    author = "Curtin, David and Gemmell, Caleb and Verhaaren, Christopher B.",
    title = "{Simulating glueball production in Nf=0 QCD}",
    eprint = "2202.12899",
    archivePrefix = "arXiv",
    primaryClass = "hep-ph",
    doi = "10.1103/PhysRevD.106.075015",
    journal = "Phys. Rev. D",
    volume = "106",
    number = "7",
    pages = "075015",
    year = "2022"
}

@article{Forshaw:2020wrq,
    author = {Forshaw, Jeffrey R. and Holguin, Jack and Pl{\"a}tzer, Simon},
    title = "{Building a consistent parton shower}",
    eprint = "2003.06400",
    archivePrefix = "arXiv",
    primaryClass = "hep-ph",
    reportNumber = "MAN/HEP/2020/002, UWTHPH-2020-8, MCnet-20-11",
    doi = "10.1007/JHEP09(2020)014",
    journal = "JHEP",
    volume = "09",
    pages = "014",
    year = "2020"
}

@article{Bauer:2022hpo,
    author = "Bauer, Christian W. and others",
    title = "{Quantum Simulation for High-Energy Physics}",
    eprint = "2204.03381",
    archivePrefix = "arXiv",
    primaryClass = "quant-ph",
    reportNumber = "UMD-PP-022-04, LA-UR-22-22100, RIKEN-iTHEMS-Report-22, RIKEN-iTHEMS-Report-22,
  FERMILAB-PUB-22-249-SQMS-T, FERMILAB-PUB-22-249-SQMS-T, IQuS@UW-21-027, MITRE-21-03848-2, FERMILAB-PUB-22-249-SQMS-T",
    doi = "10.1103/PRXQuantum.4.027001",
    journal = "PRX Quantum",
    volume = "4",
    number = "2",
    pages = "027001",
    year = "2023"
}

@article{Fang:2024ple,
    author = "Fang, Yaquan and Gao, Christina and Li, Ying-Ying and Shu, Jing and Wu, Yusheng and Xing, Hongxi and Xu, Bin and Xu, Lailin and Zhou, Chen",
    title = "{Quantum frontiers in high energy physics}",
    eprint = "2411.11294",
    archivePrefix = "arXiv",
    primaryClass = "hep-ph",
    reportNumber = "USTC-ICTS/PCFT-24-47",
    doi = "10.1007/s11433-024-2635-4",
    journal = "Sci. China Phys. Mech. Astron.",
    volume = "68",
    number = "6",
    pages = "260301",
    year = "2025"
}

@article{Yamazaki:2025bfq,
    author = "Yamazaki, Masahito",
    title = "{Quantum Wishlist: Lessons from Parton Showers}",
    eprint = "2502.18059",
    archivePrefix = "arXiv",
    primaryClass = "hep-ph",
    doi = "10.22323/1.483.0260",
    journal = "PoS",
    volume = "QCHSC24",
    pages = "260",
    year = "2025"
}

@article{Bauer:2019qxa,
    author = "Bauer, Christian W. and de Jong, Wibe A. and Nachman, Benjamin and Provasoli, Davide",
    title = "{Quantum Algorithm for High Energy Physics Simulations}",
    eprint = "1904.03196",
    archivePrefix = "arXiv",
    primaryClass = "hep-ph",
    doi = "10.1103/PhysRevLett.126.062001",
    journal = "Phys. Rev. Lett.",
    volume = "126",
    number = "6",
    pages = "062001",
    year = "2021"
}

@article{Macaluso:2021ngq,
    author = "Macaluso, Sebastian and Cranmer, Kyle",
    title = "{The Quantum Trellis: A classical algorithm for sampling the parton shower with interference effects}",
    eprint = "2112.12795",
    archivePrefix = "arXiv",
    primaryClass = "hep-ph",
    month = "12",
    year = "2021"
}

@article{Gustafson:2022dsq,
    author = {Gustafson, G{\"o}sta and Prestel, Stefan and Spannowsky, Michael and Williams, Simon},
    title = "{Collider events on a quantum computer}",
    eprint = "2207.10694",
    archivePrefix = "arXiv",
    primaryClass = "hep-ph",
    doi = "10.1007/JHEP11(2022)035",
    journal = "JHEP",
    volume = "11",
    pages = "035",
    year = "2022"
}

@article{Deliyannis:2022uyh,
    author = {Deliyannis, Plato and Sud, James and Chamaki, Diana and Webb-Mack, Zo{\"e} and Bauer, Christian W. and Nachman, Benjamin},
    title = "{Improving quantum simulation efficiency of final state radiation with dynamic quantum circuits}",
    eprint = "2203.10018",
    archivePrefix = "arXiv",
    primaryClass = "hep-ph",
    doi = "10.1103/PhysRevD.106.036007",
    journal = "Phys. Rev. D",
    volume = "106",
    number = "3",
    pages = "036007",
    year = "2022"
}

@article{Bepari:2021kwv,
    author = "Bepari, Khadeejah and Malik, Sarah and Spannowsky, Michael and Williams, Simon",
    title = "{Quantum walk approach to simulating parton showers}",
    eprint = "2109.13975",
    archivePrefix = "arXiv",
    primaryClass = "hep-ph",
    doi = "10.1103/PhysRevD.106.056002",
    journal = "Phys. Rev. D",
    volume = "106",
    number = "5",
    pages = "056002",
    year = "2022"
}

@article{Richardson:2018pvo,
    author = "Richardson, Peter and Webster, Stephen",
    title = "{Spin Correlations in Parton Shower Simulations}",
    eprint = "1807.01955",
    archivePrefix = "arXiv",
    primaryClass = "hep-ph",
    reportNumber = "IPPP/18/55, CERN-TH-2018-154, IPPP-18-55, MCNET-18-12",
    doi = "10.1140/epjc/s10052-019-7429-5",
    journal = "Eur. Phys. J. C",
    volume = "80",
    number = "2",
    pages = "83",
    year = "2020"
}

@article{Hamilton:2021dyz,
    author = "Hamilton, Keith and Karlberg, Alexander and Salam, Gavin P. and Scyboz, Ludovic and Verheyen, Rob",
    title = "{Soft spin correlations in final-state parton showers}",
    eprint = "2111.01161",
    archivePrefix = "arXiv",
    primaryClass = "hep-ph",
    doi = "10.1007/JHEP03(2022)193",
    journal = "JHEP",
    volume = "03",
    pages = "193",
    year = "2022"
}

@article{CMS:2025awh,
    collaboration = "CMS",
    title = "{Measurement of gluon spin effect in parton splittings}",
    journal = "CMS-PAS-SMP-25-006",
    year = "2025"
}

@article{Basham:1978bw,
    author = "Basham, C. Louis and Brown, Lowell S. and Ellis, Stephen D. and Love, Sherwin T.",
    title = "{Energy Correlations in electron - Positron Annihilation: Testing QCD}",
    reportNumber = "RLO-1388-759",
    doi = "10.1103/PhysRevLett.41.1585",
    journal = "Phys. Rev. Lett.",
    volume = "41",
    pages = "1585",
    year = "1978"
}

@article{Basham:1978zq,
    author = "Basham, C. L. and Brown, L. S. and Ellis, S. D. and Love, S. T.",
    title = "{Energy Correlations in electron-Positron Annihilation in Quantum Chromodynamics: Asymptotically Free Perturbation Theory}",
    reportNumber = "RLO-1388-761",
    doi = "10.1103/PhysRevD.19.2018",
    journal = "Phys. Rev. D",
    volume = "19",
    pages = "2018",
    year = "1979"
}

@article{Buschmann:2015awa,
    author = "Buschmann, Malte and Kopp, Joachim and Liu, Jia and Machado, Pedro A. N.",
    title = "{Lepton Jets from Radiating Dark Matter}",
    eprint = "1505.07459",
    archivePrefix = "arXiv",
    primaryClass = "hep-ph",
    reportNumber = "MITP-15-036, IFT-UAM-CSIC-15-047, FTUAM-15-13",
    doi = "10.1007/JHEP07(2015)045",
    journal = "JHEP",
    volume = "07",
    pages = "045",
    year = "2015"
}

@article{Virtanen_2020,
   title="{SciPy 1.0: fundamental algorithms for scientific computing in Python}",
   volume="17",
   journal="Nature Methods",
   doi = "10.1038/s41592-019-0686-2",
   author= "Virtanen, Pauli and Gommers, Ralf and Oliphant, Travis E. and Haberland, Matt and Reddy, Tyler and Cournapeau, David and Burovski, Evgeni and Peterson, Pearu and Weckesser, Warren and Bright, Jonathan and van der Walt, Stéfan J. and Brett, Matthew and Wilson, Joshua and Millman, K. Jarrod and Mayorov, Nikolay and Nelson, Andrew R. J. and Jones, Eric and Kern, Robert and Larson, Eric and Carey, C J and Polat, İlhan and Feng, Yu and Moore, Eric W. and VanderPlas, Jake and Laxalde, Denis and Perktold, Josef and Cimrman, Robert and Henriksen, Ian and Quintero, E. A. and Harris, Charles R. and Archibald, Anne M. and Ribeiro, Antônio H. and Pedregosa, Fabian and van Mulbregt, Paul and Vijaykumar, Aditya and Bardelli, Alessandro Pietro and Rothberg, Alex and Hilboll, Andreas and Kloeckner, Andreas and Scopatz, Anthony and Lee, Antony and Rokem, Ariel and Woods, C. Nathan and Fulton, Chad and Masson, Charles and Häggström, Christian and Fitzgerald, Clark and Nicholson, David A. and Hagen, David R. and Pasechnik, Dmitrii V. and Olivetti, Emanuele and Martin, Eric and Wieser, Eric and Silva, Fabrice and Lenders, Felix and Wilhelm, Florian and Young, G. and Price, Gavin A. and Ingold, Gert-Ludwig and Allen, Gregory E. and Lee, Gregory R. and Audren, Hervé and Probst, Irvin and Dietrich, Jörg P. and Silterra, Jacob and Webber, James T and Slavič, Janko and Nothman, Joel and Buchner, Johannes and Kulick, Johannes and Schönberger, Johannes L. and de Miranda Cardoso, José Vinícius and Reimer, Joscha and Harrington, Joseph and Rodríguez, Juan Luis Cano and Nunez-Iglesias, Juan and Kuczynski, Justin and Tritz, Kevin and Thoma, Martin and Newville, Matthew and Kümmerer, Matthias and Bolingbroke, Maximilian and Tartre, Michael and Pak, Mikhail and Smith, Nathaniel J. and Nowaczyk, Nikolai and Shebanov, Nikolay and Pavlyk, Oleksandr and Brodtkorb, Per A. and Lee, Perry and McGibbon, Robert T. and Feldbauer, Roman and Lewis, Sam and Tygier, Sam and Sievert, Scott and Vigna, Sebastiano and Peterson, Stefan and More, Surhud and Pudlik, Tadeusz and Oshima, Takuya and Pingel, Thomas J. and Robitaille, Thomas P. and Spura, Thomas and Jones, Thouis R. and Cera, Tim and Leslie, Tim and Zito, Tiziano and Krauss, Tom and Upadhyay, Utkarsh and Halchenko, Yaroslav O. and Vázquez-Baeza, Yoshiki",
   year="2020",
   month="2", pages="261"}

@article{Bassetto:1983mvz,
    author = "Bassetto, A. and Ciafaloni, M. and Marchesini, G.",
    title = "{Jet Structure and Infrared Sensitive Quantities in Perturbative QCD}",
    doi = "10.1016/0370-1573(83)90083-2",
    journal = "Phys. Rept.",
    volume = "100",
    pages = "201--272",
    year = "1983"
}

@article{Amati:1980ch,
    author = "Amati, D. and Bassetto, A. and Ciafaloni, M. and Marchesini, G. and Veneziano, G.",
    title = "{A Treatment of Hard Processes Sensitive to the Infrared Structure of QCD}",
    reportNumber = "CERN-TH-2831",
    doi = "10.1016/0550-3213(80)90012-7",
    journal = "Nucl. Phys. B",
    volume = "173",
    pages = "429--455",
    year = "1980"
}

@article{Bahr:2008pv,
    author = "Bahr, M. and others",
    title = "{Herwig++ Physics and Manual}",
    eprint = "0803.0883",
    archivePrefix = "arXiv",
    primaryClass = "hep-ph",
    reportNumber = "CERN-PH-TH-2008-038, CAVENDISH-HEP-08-03, KA-TP-05-2008, DCPT-08-22, IPPP-08-11, CP3-08-05",
    doi = "10.1140/epjc/s10052-008-0798-9",
    journal = "Eur. Phys. J. C",
    volume = "58",
    pages = "639--707",
    year = "2008"
}

@article{Bloch:2026rrw,
    author = "Bloch, Itay M. and Carena, Marcela and Chen, Yifan and Li, Xinran and Li, Ying-Ying and Shu, Jing",
    title = "{Particle Physics Driven by Quantum Technology - Quantum Simulations and Quantum Sensing}",
    eprint = "2609.20500",
    archivePrefix = "arXiv",
    primaryClass = "hep-ph",
    reportNumber = "FERMILAB-PUB-26-0274-T",
    month = "9",
    year = "2026"
}

@article{Li:2026rfq,
    author = "Li, Yin and Wang, Jian Bin and Xie, Jia Qi and Xu, Kai Rong and Ye, Rui Han and Huang, Zi Huan and Liu, Bingxuan",
    title = "{Reconstructing the invisible fraction of semivisible jets in ISR-boosted events via neural network regression}",
    eprint = "2604.20456",
    archivePrefix = "arXiv",
    primaryClass = "hep-ph",
    doi = "10.1103/hbxx-5f3k",
    journal = "Phys. Rev. D",
    volume = "114",
    number = "5",
    pages = "056017",
    year = "2026"
}

@article{Batz:2023zef,
    author = "Batz, Austin and Cohen, Timothy and Curtin, David and Gemmell, Caleb and Kribs, Graham D.",
    title = "{Dark sector glueballs at the LHC}",
    eprint = "2310.13731",
    archivePrefix = "arXiv",
    primaryClass = "hep-ph",
    reportNumber = "CERN-TH-2023-194",
    doi = "10.1007/JHEP04(2024)070",
    journal = "JHEP",
    volume = "04",
    pages = "070",
    year = "2024"
}

@article{Hofman:2008ar,
    author = "Hofman, Diego M. and Maldacena, Juan",
    title = "{Conformal collider physics: Energy and charge correlations}",
    eprint = "0803.1467",
    archivePrefix = "arXiv",
    primaryClass = "hep-th",
    doi = "10.1088/1126-6708/2008/05/012",
    journal = "JHEP",
    volume = "05",
    pages = "012",
    year = "2008"
}

@article{Dixon:2019uzg,
    author = "Dixon, Lance J. and Moult, Ian and Zhu, Hua Xing",
    title = "{Collinear limit of the energy-energy correlator}",
    eprint = "1905.01310",
    archivePrefix = "arXiv",
    primaryClass = "hep-ph",
    reportNumber = "SLAC-PUB-17427, SLAC--PUB--17427",
    doi = "10.1103/PhysRevD.100.014009",
    journal = "Phys. Rev. D",
    volume = "100",
    number = "1",
    pages = "014009",
    year = "2019"
}

@article{Korchemsky:2019nzm,
    author = "Korchemsky, G. P.",
    title = "{Energy correlations in the end-point region}",
    eprint = "1905.01444",
    archivePrefix = "arXiv",
    primaryClass = "hep-th",
    reportNumber = "IPhT-T19/041",
    doi = "10.1007/JHEP01(2020)008",
    journal = "JHEP",
    volume = "01",
    pages = "008",
    year = "2020"
}

\end{document}